\documentclass[
 reprint,
superscriptaddress,
 amsmath,amssymb,
 aps,
prl,
]{revtex4-1}

\newcommand{\ion}[2]{\mbox{$^{#2}$#1$^+$}}
\newcommand{\Yb}[1]{\ion{Yb}{#1}}

\newcommand{\unit}[1]{\,\mbox{#1}}

\newcommand{\us}{\unit{$\mu$s}}

\newcommand{\Omuw}{\mbox{$\Omega_{\rm \mu w}$}}
\newcommand{\Omrf}{\mbox{$\Omega_{\rm rf}$}}

\newcommand{\bla}[1]{\left( #1 \right)}
\newcommand{\blb}[1]{\left[ #1 \right]}
\newcommand{\blc}[1]{\left\lbrace #1 \right\rbrace}

\usepackage{graphicx}
\usepackage{tabularx}
\usepackage{amsmath}

\usepackage{graphicx}
\usepackage{dcolumn}
\usepackage{bm}
\usepackage{lineno}

\newcommand{\bra}[1]{\ensuremath{\left\langle#1\right|}}
\newcommand{\ket}[1]{\ensuremath{\left|#1\right\rangle}}

\DeclareGraphicsExtensions{.pdf,.png,.jpg,.eps,.tiff}

\begin{document}

\author{S. Weidt}
\affiliation{Department of Physics and Astronomy, University of Sussex, Brighton, BN1 9QH, UK}
\author{J. Randall}
\affiliation{Department of Physics and Astronomy, University of Sussex, Brighton, BN1 9QH, UK}
\affiliation{QOLS, Blackett Laboratory, Imperial College London, London, SW7 2BW, UK}
\author{S. C. Webster}
\affiliation{Department of Physics and Astronomy, University of Sussex, Brighton, BN1 9QH, UK}
\author{K. Lake}
\affiliation{Department of Physics and Astronomy, University of Sussex, Brighton, BN1 9QH, UK}
\author{A. E. Webb}
\affiliation{Department of Physics and Astronomy, University of Sussex, Brighton, BN1 9QH, UK}
\author{I. Cohen}
\affiliation{Racah Institute of Physics, The Hebrew University of Jerusalem, Jerusalem 91904, Givat Ram, Israel}
\author{T. Navickas}
\affiliation{Department of Physics and Astronomy, University of Sussex, Brighton, BN1 9QH, UK}
\author{B. Lekitsch}
\affiliation{Department of Physics and Astronomy, University of Sussex, Brighton, BN1 9QH, UK}
\author{A. Retzker}
\affiliation{Racah Institute of Physics, The Hebrew University of Jerusalem, Jerusalem 91904, Givat Ram, Israel}
\author{W. K. Hensinger}
\email{W.K.Hensinger@sussex.ac.uk}
\affiliation{Department of Physics and Astronomy, University of Sussex, Brighton, BN1 9QH, UK}
\title{Trapped-ion quantum logic with global radiation fields}

\begin{abstract}
Trapped ions are a promising tool for building a large-scale quantum computer. However, the number of required radiation fields for the realisation of quantum gates in any proposed ion-based architecture scales with the number of ions within the quantum computer, posing a major obstacle when imagining a device with millions of ions. Here we present a fundamentally different concept for trapped-ion quantum computing where this detrimental scaling entirely vanishes, replacing millions of radiation fields with only a handful of fields. The method is based on individually controlled voltages applied to each logic gate location to facilitate the actual gate operation analogous to a traditional transistor architecture within a classical computer processor. To demonstrate the key principle of this approach we implement a versatile quantum gate method based on long-wavelength radiation and use this method to generate a maximally entangled state of two quantum engineered clock-qubits with fidelity 0.985(12). This quantum gate also constitutes a simple-to-implement tool for quantum metrology, sensing and simulation.
\end{abstract}

\let\originalnewpage\newpage
\let\newpage\relax
\maketitle
\let\newpage\originalnewpage

The control of the internal and external degrees of freedom of trapped ions using laser light has allowed unprecedented advances in the creation of multi-particle entangled states \cite{Blatt1,Haffner3,Sackett,Leibfried3}, quantum simulation \cite{Blatt,Friedenauer,Kim,Lanyon,Johanning,Schneider1}, frequency standards \cite{Ludlow}, quantum sensing \cite{Kotler,Pruttivarasin,Baumgart} and quantum logic \cite{Haffner2,Leibfried2}. A major goal is now to construct a large-scale quantum computer by scaling current systems up to a significantly larger number of ions \cite{Cirac,Kielpinski,Monroe4}. The circuit-model approach for quantum information processing requires the realization of single qubit gates and a two-qubit entanglement operation \cite{Nielsen}. The use of laser light for the implementation of these quantum logic operations has been extremely successful, with gate fidelities in the fault-tolerant regime having been achieved for single \cite{Akerman, Ballance} as well as two-qubit gates \cite{Gaebler, Ballance}.

Instead of using laser light it is also possible to use long-wavelength radiation in the microwave and RF regime to implement quantum logic. Such fields are comparably simple to generate and highly stable and have already been used to implement single-qubit gates with errors of only $10^{-6}$, far surpassing fault-tolerant thresholds \cite{Harty}. Free-running long-wavelength radiation on its own is however not sufficient for the implementation of multi-qubit gates, as it only weakly drives the ions' motion due to the vanishingly small Lamb-Dicke parameter. This drawback was first addressed in the seminal work by Mintert and Wunderlich in 2001 who showed that combining a static magnetic field gradient with externally applied long-wavelength radiation creates a sizable effective Lamb-Dicke parameter \cite{Mintert1}. More recently, Ospelkaus et al. proposed using the oscillating magnetic field gradients experienced by an ion trapped in the near-field of a microwave waveguide to perform multi-qubit gates \cite{Ospelkaus1}. This scheme was subsequently used to perform the first microwave-based two-qubit gate by Ospelkaus et al. \cite{Ospelkaus}. The scheme requires ions to be trapped close to a surface incorporating the microwave waveguide and therefore the effects of motional heating must be more carefully considered. When scaling this approach, especially considering complicated electrode geometries such as X-junctions, relevant individual microwave impedance matching for each gate zone across the whole architecture must be assured. Addressing of individual ions would typically require the use of destructive interference incorporating all microwave fields applied within the range of the ion or other sophisticated methods \cite{Warring, Craik}.

\begin{figure*}
\centering
\includegraphics[width=1\textwidth]{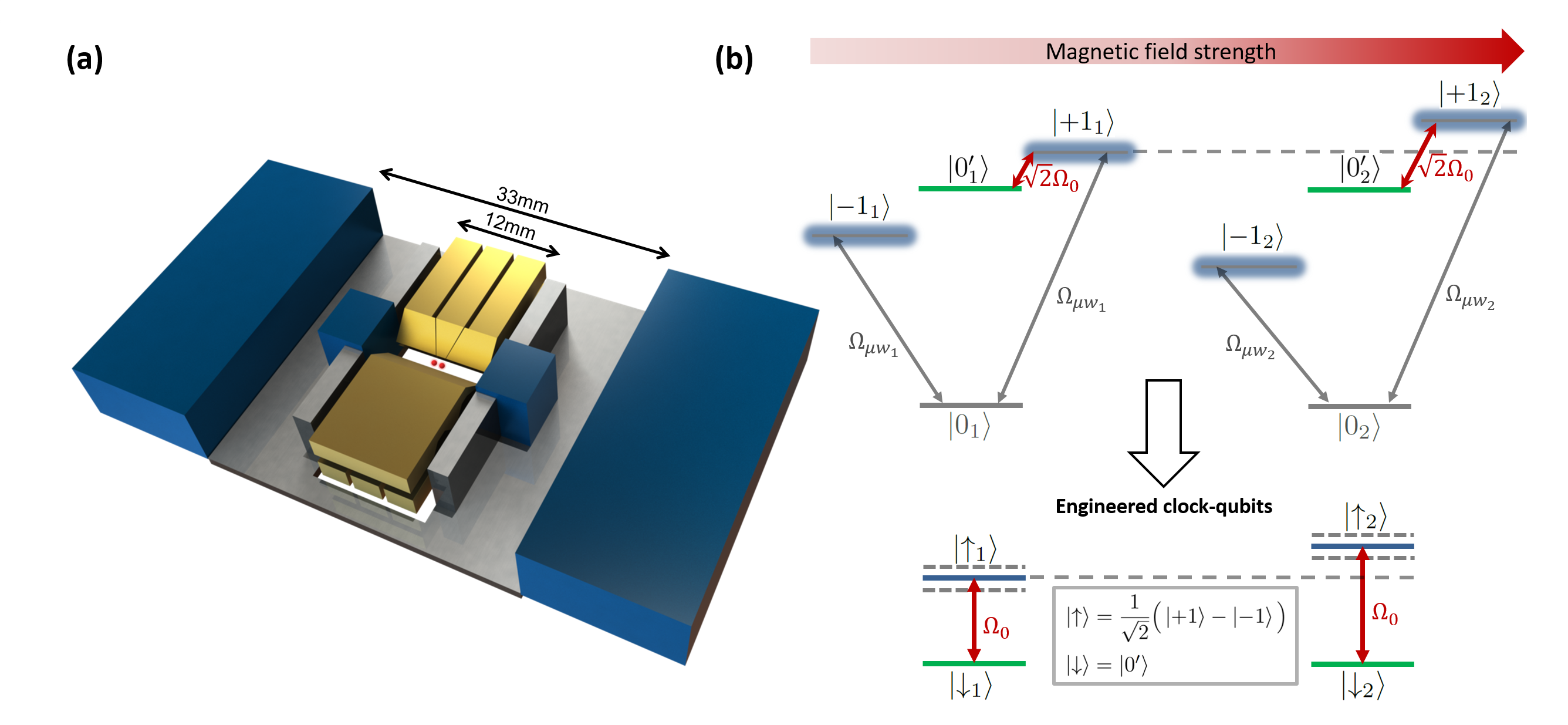}
\caption{(a) Schematic of our linear Paul trap (yellow) fitted with four permanent magnets (blue), arranged to create a strong magnetic field gradient along the trap axis. (b) Illustration of the $^{2}$S$_{1/2}$ ground-state hyperfine manifold of two $^{171}$Yb$^{+}$ ions, each being driven by two resonant microwave fields near 12.6 GHz with slightly unequal Rabi frequencies denoted by $\Omega_{\mu \text{w}_{1}}$ and $\Omega_{\mu \text{w}_{2}}$ (Supplementary Methods). The engineered clock qubit is formed of $\ket{\uparrow}=(\ket{+1}-\ket{-1})/\sqrt{2}$ and $\ket{\downarrow}=\ket{0'}$ which can be manipulated using an RF field coupling $\ket{0'}$ and $\ket{+1}$ with Rabi frequency $\sqrt{2}\Omega_{0}$.}
\label{experiment_setup}
\end{figure*}

The approach of using a static magnetic field gradient in conjunction with externally applied long-wavelength radiation is not subject to the above constraints (of course the effects of motional heating should still be considered) and has also been used to implement a two-qubit gate between nearest as well as non-nearest neighbour ions \cite{Khromova1}. In stark contrast to the work presented in this manuscript, the first demonstration of using a static magnetic field gradient to implement a two-qubit gate made use of an `undriven' magnetic gradient induced coupling. However, in this scheme the dominant source of noise is ambient magnetic field fluctuations as naturally occurring states with different magnetic moments must be used, ruling out the use of a so-called clock qubit. A promising approach to circumvent this drawback is to use `dressed states' \cite{Timoney,Tan,Webster} where one can quantum engineer an effective clock qubit which is highly protected from magnetic field fluctuations while maintaining a strong sensitivity to a static magnetic field gradient. They have already been used in single qubit operations \cite{Timoney,Webster} and to cool an ion close to its ground state of motion \cite{Weidt} and their use to implement a two qubit gate would constitute a significant breakthrough for quantum computing with long-wavelength radiation.

Despite these successes, scaling these laser or long-wavelength radiation based operations to a much larger number of ions constitutes a tremendous challenge. This becomes particularly obvious when considering that a large-scale universal quantum computer, say of the size large enough to break RSA encoding, would require millions or even billions of qubits \cite{Monroe4, Lekitsch}. Gate operations need to be carried out in parallel for the quantum computer to work. The implication of this is that a large-scale quantum computer may require millions of spatially separated `gate zones' where quantum gates are executed. This results in the requirement of utilizing millions of laser or long-wavelength radiation fields for the implementation of quantum gates when considering all previous proposals to build a large-scale trapped-ion quantum computer \cite{Cirac, Kielpinski, Monroe3, Monroe4}. This detrimental scaling between the number of ions and the required number of radiation fields constitutes a significant obstacle to scaling to the desired large system sizes.  

In this work we remove this obstacle. We present a new concept for trapped-ion quantum computing where parallel quantum gate operations in arbitrarily many selected gate zones can be executed using individually controlled voltages applied to each gate zone. Instead of millions of laser or long-wavelength radiation fields this remarkably simple approach only requires a handful of global radiation fields where the number of radiation fields only depends on the number of different types of quantum gates to be executed in parallel. This then provides a simple and powerful concept for quantum computing which forms the core element within a wider engineering blueprint to build a large-scale microwave-based trapped-ion quantum computer \cite{Lekitsch}. A key element of our approach is the use of qubits which feature a widely tunable transition frequency while maintaining its protected nature with respect to ambient magnetic field fluctuations. Quantum engineered clock qubits meet this requirement and therefore constitute an ideal system for this purpose. We demonstrate the key element of our approach by generating entanglement between microwave-based quantum-engineered clock qubits in a M{\o}lmer-S{\o}rensen-type interaction utilizing long-wavelength radiation and a static magnetic field gradient \cite{Sorensen2, Timoney}. 
 
The two-qubit gate is performed on two $^{171}$Yb$^{+}$ ions in a Paul trap with an ion-electrode distance of 310 $\mu$m \cite{McLoughlin2}. We place permanent magnets close to the ion trap with an ion-to-nearest-magnet distance of approximately 6 mm as shown in Fig. 1. This provides a static magnetic field gradient of 23.6(3) T/m which is approximately constant across the ion string \cite{Lake}. We slightly displace the ions from the magnetic field nil, which lifts the degeneracy of the $^{2}$S$_{1/2}$ F=1 manifolds by 12.0 MHz and 14.8 MHz for ions 1 and 2 respectively, and defines the internal-state quantisation axis to lie along the trap axis. Laser light near-resonant with the $^{2}$S$_{1/2}\leftrightarrow$ $^{2}$P$_{1/2}$ transition is used for Doppler laser cooling and for initial state preparation as well as state detection. State-dependent fluorescence is collected on a photo-multiplier tube, and the fluorescence measurements are normalised to remove preparation and detection errors (Supplementary Methods).

\begin{figure}[tb]
\centering
\includegraphics[width=0.45\textwidth]{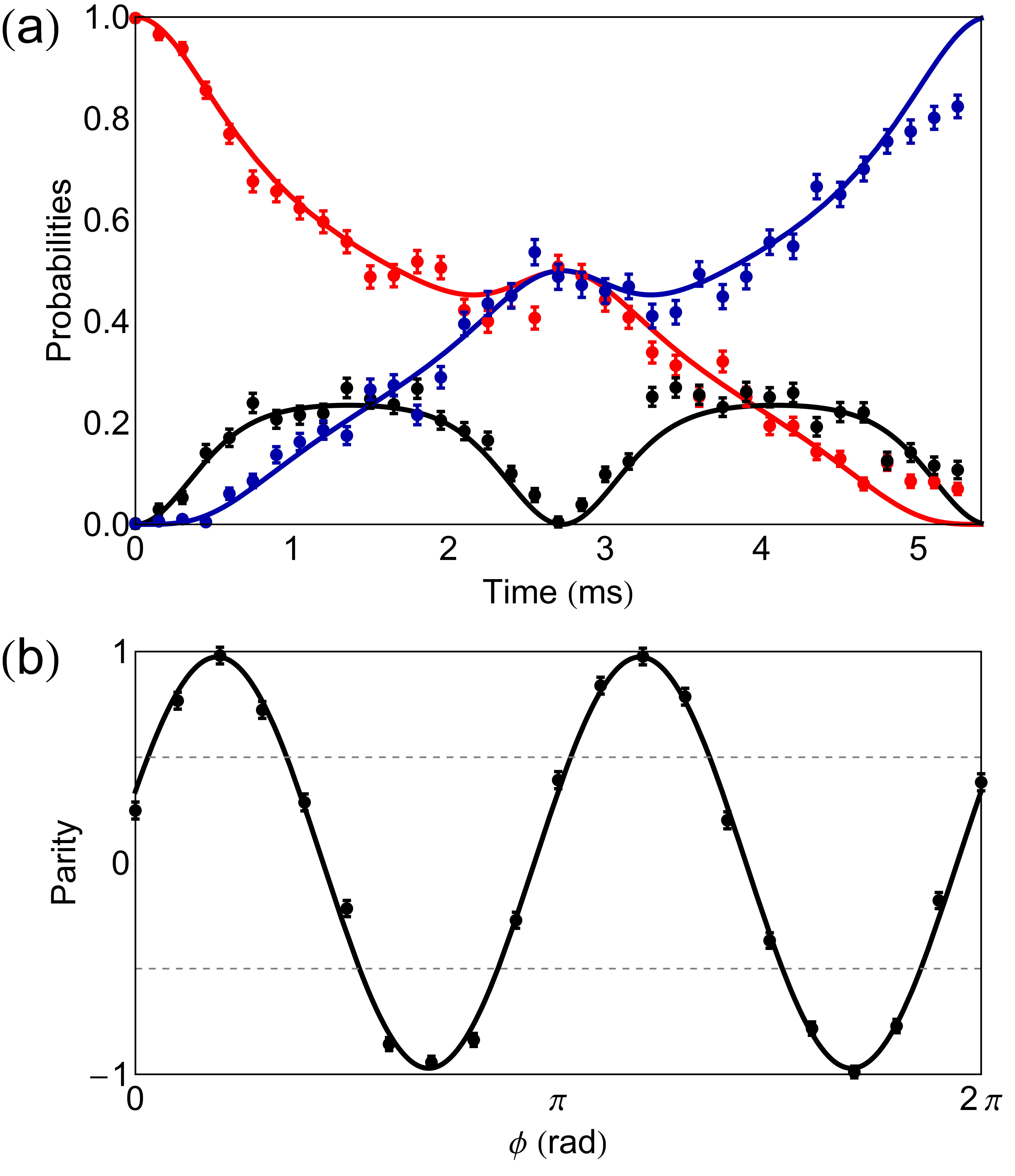}
\caption{(a) Populations $P(\uparrow\uparrow)$ (red), $P(\downarrow\downarrow)$ (blue) and $P(\uparrow\downarrow)+P(\downarrow\uparrow)$ (black) after preparing the ion spins in the state $\ket{\downarrow\downarrow}$ and applying the M{\o}lmer-S{\o}rensen fields for a variable time \textit{t}. A maximally entangled state is formed at time $t_g = 2.7$ ms. Each data point is the average of 500 measurements and the solid lines are the predicted theoretical curves. (b) Parity $\Pi = P(\uparrow\uparrow)+P(\downarrow\downarrow)-P(\uparrow\downarrow)-P(\downarrow\uparrow)$ after applying the M{\o}lmer-S{\o}rensen interaction for a time $t_g$, followed by a $\pi/2$ pulse on each ion with variable phase $\phi$. The signal oscillates as $\cos(2\phi)$, with an amplitude $A$ that indicates the magnitude of the off-diagonal density matrix elements $|\rho_{\downarrow\downarrow,\uparrow\uparrow}|$ \cite{Sackett}. Each data point is the average of 800 measurements and the black line is a fit to the data.}
\label{gate_populations}
\end{figure}

To globally broadcast the required long-wavelength radiation we only require a standard off-the-shelf microwave horn and a 3-turn rf-emitting copper coil placed outside the ultra-high vacuum environment. We note that in a large-scale architecture our approach utilizes submerged static currents incorporated into the microfabricated chip traps to give rise to the required static magnetic field gradients. The ion-surface distance requirement in this case is not very stringent. Simulations show magnetic field gradients in excess of 150 T/m with an ion-electrode distance of approximately 150 $\mu m$ can be achieved, using realistic values of applied current that have already been applied to an ion trapping chip of this type \cite{Lekitsch}. Such a relatively large ion-electrode distance minimizes motional decoherence due to charge fluctuations from the electrode surface. 

Instead of using a naturally occurring magnetic field sensitive qubit we quantum engineer a tunable highly noise-resilient `clock-like' qubit by first addressing each ion with a pair of microwave fields coupling the $^2$S$_{1/2}, F=0 \equiv\ket{0}$ with the $^2$S$_{1/2}, F=1, m_F=+1 \equiv\ket{+1}$ and $^2$S$_{1/2}, F=1, m_{F}=-1 \equiv\ket{-1}$ (Fig. 1). In the appropriate interaction picture this results in three dressed-states, including the well-protected state $\ket{\uparrow}=(\ket{+1}-\ket{-1})/\sqrt{2}$ \cite{Timoney}. We combine this state with the intrinsically well-protected state $^{2}S_{1/2} F = 1, m_{F} = 0 \equiv \ket{\downarrow}$ to obtain a quantum-engineered clock qubit  $\{$$\ket{\downarrow}$ , $\ket{\downarrow}$$\}$ (Supplementary Methods). Unlike a standard clock transition, which has a fixed transition frequency, the qubit transition frequency is tunable using a magnetic-field, enabling individual qubit addressing with global radiation fields. This is a critical feature when viewed within the context of the concept for a trapped-ion quantum computer outlined below. We prepare and detect the engineered clock qubit using the method developed by Randall et al. \cite{Randall2}. Arbitrary single qubit gates between $\ket{\downarrow}$ and $\ket{\uparrow}$ are implemented using an RF field resonant with the $\ket{\downarrow}\leftrightarrow$ $\ket{+1}$ transition \cite{Webster}. The degeneracy in frequency between this and the $\ket{\downarrow}\leftrightarrow$ $\ket{-1}$ transition is lifted by the second-order Zeeman shift. Using a Ramsey type experiment we measure the coherence time of this qubit to be 650 ms, significantly longer than the $\approx 1$ ms coherence time of the bare state qubits that have so far been used for two-qubit gates with a static magnetic field gradient.

We create a maximally entangled state using a M{\o}lmer-S{\o}rensen type gate. The application of this gate to our qubit has been investigated in detail theoretically \cite{Timoney,Gatis} and forms the basis of our experimental implementation. We implement the gate on the axial stretch mode with a frequency of $\nu_s=\sqrt{3}\nu_z = 2\pi \times 459.34(1)$ kHz, where $\nu_z$ is the axial centre-of-mass mode frequency, giving an effective Lamb-Dicke parameter \cite{Mintert1} $\eta_\text{eff} = z_0 \mu_B \partial_zB/\sqrt{2}\hbar \nu_s = 0.0041$, where $z_0 = \sqrt{\hbar/2 m \nu_s}$. This mode is sideband cooled to $\bar{n} = 0.14(3)$ using a variant of the scheme described in ref. \cite{Weidt} (Supplementary Methods) before the internal states are prepared in the state $\ket{\downarrow\downarrow}$. A pair of RF fields is then applied to each ion with frequencies close to the red and blue sidebands (carrier Rabi frequency $\Omega_0 = 2\pi \times 45.4$ kHz). The frequencies are set to be symmetric about the carrier frequency, corresponding to detunings $\pm \nu_s \pm \delta$. The gate detuning $\delta$ is set to $\delta = 2\eta_\text{eff}\Omega_0 = 2\pi \times 370$ Hz in order that at time $t_g = 2\pi/\delta = 2.7$ ms, the ions are ideally prepared in a maximally entangled spin state $\ket{\Psi_{\phi_{0}}} = (\ket{\uparrow\uparrow} + e^{i\phi_{0}} \ket{\downarrow\downarrow})/\sqrt{2}$ (Supplementary Methods). Fig. 2 (a) shows the evolution of the spin state populations as a function of time. To measure the coherence of the entangled state, a carrier $\pi/2$ pulse is applied to each ion after the gate pulse. Fig. 2 (b) shows the parity $\Pi = P(\uparrow\uparrow)+P(\downarrow\downarrow)-P(\uparrow\downarrow)-P(\downarrow\uparrow)$ as a function of the phase $\phi$ of the $\pi/2$ pulse (Supplementary Methods). The amplitude of the parity oscillation (a) along with the populations at $t_g$ allows the fidelity of the obtained density matrix $\hat{\rho}$ with respect to the ideal outcome $\ket{\Psi_{\phi_{0}}}$ to be calculated using $\mathcal{F} = \bra{\Psi_{\phi_{0}}}\hat{\rho}\ket{\Psi_{\phi_{0}}} = [P(\uparrow\uparrow) + P(\downarrow\downarrow)]/2 + A/2$ \cite{Sackett}. We measure the populations at $t_g$ to be $P(\uparrow\uparrow) + P(\downarrow\downarrow) = 0.997(8)$ and a fit to the parity scan shown in Fig. 2 (b) gives an amplitude of $A = 0.972(17)$. From this we extract a Bell state fidelity of $\mathcal{F} = 0.985(12)$.

The most significant contributions to the infidelity stem from heating of the vibrational mode of motion ($1\times 10^{-2}$) used during the gate operation and depolarisation of the qubit ($3\times 10^{-3}$). Both sources of error can be significantly reduced by increasing the gate speed using a larger static magnetic field gradient and by increasing $\Omega_0$. The depolarisation error can be further reduced by improving our microwave setup as a result of which a coherence time of seconds should be achievable as already demonstrated by Baumgart et al. \cite{Baumgart}. Additional small sources of infidelity are discussed in the Supplementary Methods. 

Achieving gate fidelities that would enable fault-tolerant operation using long-wavelength radiation can be realized either by the use of ion trap microchips or by a slight modification of our setup. By reducing the ion-to-nearest-magnet distance in a modified trap design to 2.4 mm, a magnetic field gradient of 150 T/m would result. This gives a large increase of the motional coupling strength, enabling a significant reduction of the error terms. Following a full numerical simulation of the system, a fidelity far in the fault-tolerant regime would result using already demonstrated parameters (Supplementary Methods).

\begin{figure}
\centering
\includegraphics[width=0.5\textwidth]{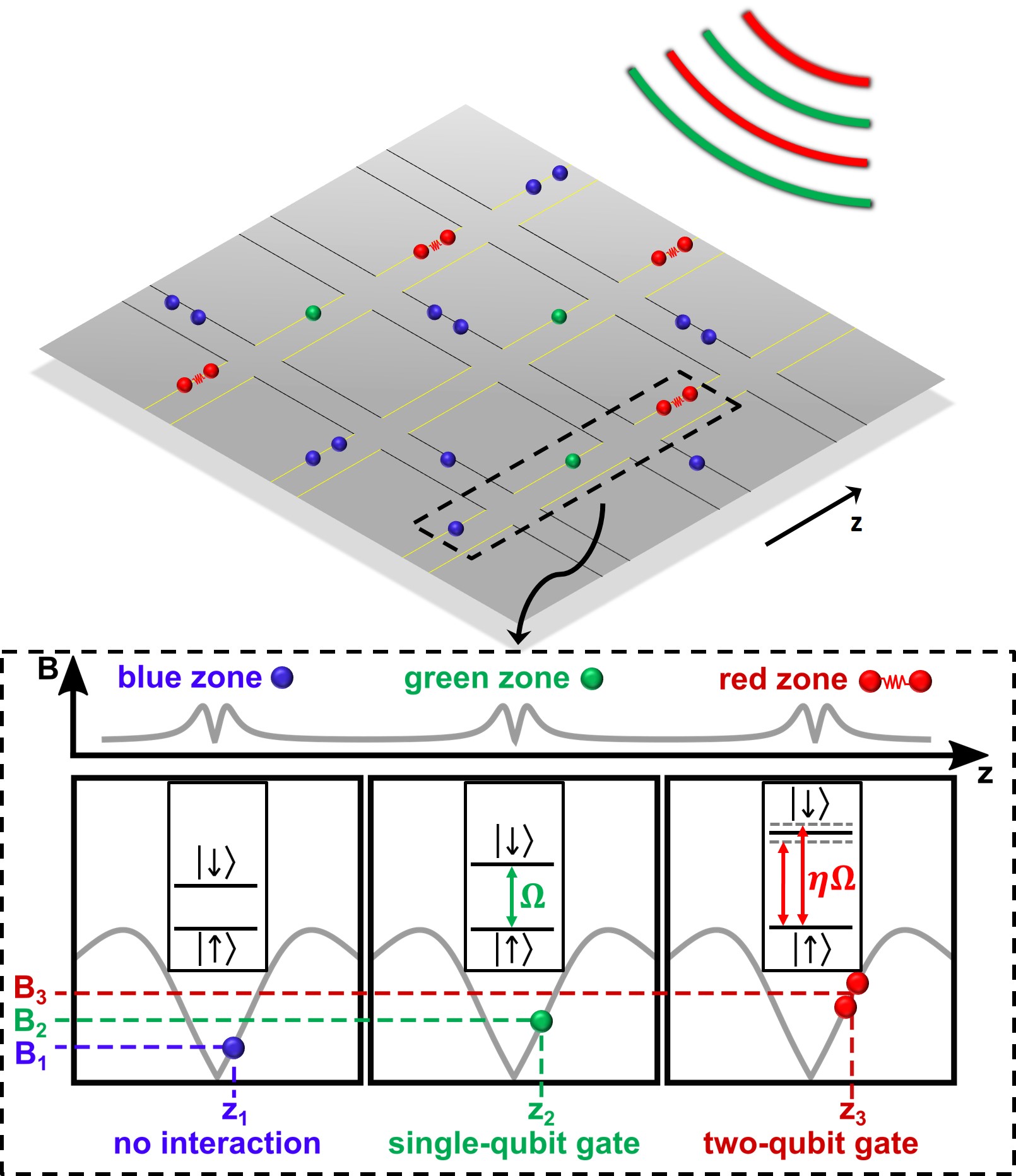}
\caption{Ions are confined in a two-dimensional X-junction surface trap architecture. Local DC electrodes are used to shift the centre of the trapping potentials in the magnetic field gradient in order to tune a particular zone in resonance with a particular set of microwave and RF fields (illustrated in the dashed box). The ion displacements in the green (red) zones tune the respective ions into resonance with the global fields to realize single- (two-) qubit gates while no shift is applied to the blue zones, making all globally applied fields off-resonant for ions located in these zones. Current-carrying wires (not shown for clarity) located below each gate zone (indicated by yellow lines) create a static magnetic field gradient local within each gate zone.}
\label{scaling}
\end{figure}

We now describe how the gate method explained above gives rise to a highly efficient approach to quantum computing with trapped ions. In previously envisioned trapped-ion quantum computing architectures the number of radiation fields required for quantum gate implementation is strongly correlated with the number of ions used \cite{Cirac,Kielpinski,Monroe4}. This constitutes a substantial challenge in the construction of a large-scale quantum computer, which may require the manipulation of millions or billions of ions. We will now outline an approach that completely removes this undesirable correlation where millions of laser or long-wavelength radiation fields are replaced with only a handful of long-wavelength radiation fields. 

Ions are located in individual gate zones that are contained within an array of X-junctions as part of a microfabricated ion trap architecture (see Fig. \ref{scaling}). Currents applied locally to each gate zone create magnetic field gradients of 150 T/m, to be used for entanglement generation. In order to select any arbitrary set of gate zones for single- or two-qubit gate execution, one simply shifts the position of the ion(s) within these zones axially with respect to the magnetic field gradient by an appropriate amount using local DC electrodes already used for ion transport within the ion trap array. In a magnetic field gradient, such shifts in the ion positions result in a variation of the local offset magnetic field. The transition frequency of the quantum engineered clock-qubit used in this work can be changed using such offset magnetic fields. This provides the ability to tune the quantum engineered clock qubit into and out of resonance with globally applied long-wavelength radiation fields. Therefore, ions in any arbitrary zone can be tuned into resonance with a set of globally applied microwave and RF fields (of the sort used to implement the two-qubit gate presented in this manuscript), providing parallel execution of gates in relevant zones while all other zones on the architecture remain off-resonant. Alternatively, instead of using the displacement of the ions to change the offset magnetic field, an offset magnetic field could be generated using additional local magnetic field coils located under each gate zone. Microwave horns and antennas located outside the vacuum system broadcast the required set of microwave and RF fields over the entire microchip or quantum computer architecture. Quantum operations are then applied in parallel to arbitrarily many sets of qubits with negligible crosstalk (Supplementary Methods) using a small number of offset magnetic fields and associated sets of global microwave and RF fields, as shown in Fig. 3. This approach, particularly when viewed within the context of a blueprint providing necessary technical details of a realistic device \cite{Lekitsch} may provide some foundation for the assertion that the construction of a practical trapped-ion quantum computer is now within reach of current technology.

\subsection*{Acknowledgements}
We thank Klaus M\o{}lmer for helpful discussions and Eamon Standing for performing relevant magnetic field simulations. This work is supported by the U.K. Engineering and Physical Sciences Research Council [EP/G007276/1, the UK Quantum Technology hub for Networked Quantum Information Technologies (EP/M013243/1), the UK Quantum Technology hub for Sensors and Metrology(EP/M013243/1)], the European Commissions Seventh Framework Programme (FP7/2007-2013) under Grant Agreement No. 270843 (iQIT), the Army Research Laboratory under Cooperative Agreement No. W911NF-12-2-0072, the US Army Research Office Contract No. W911NF-14-2-0106, and the University of Sussex. The views and conclusions contained in this document are those of the authors and should not be interpreted as representing the official policies, either expressed or implied, of the Army Research Laboratory or the U.S. Government. The U.S. Government is authorized to reproduce and distribute reprints for Government purposes notwithstanding any copyright notation herein.

\section*{Supplementary material}

\subsection*{Atomic structure}
As described in the main text, the gate is performed between two trapped \Yb{171} ions which sit in different magnetic fields due to a magnetic-field gradient. The Hamiltonian consists of three terms $H=H_{\rm int}+H_{\rm ext}+H_{\rm couple}$, where $H_{\rm int}$ describes the four internal states (\ket{0}, \ket{0'}, \ket{-1}, and \ket{+1}) of each atom, $H_{\rm ext}$ describes the axial stretch mode of the ion pair, and $H_{\rm couple}$ describes the coupling between the internal and external degrees of freedom due to the field gradient.

\begin{eqnarray}
& & \hspace{1.1em}\begin{split}H_{\rm int} = \sum\limits_{i=1,2}&-\omega_i^{0}\ket{0}_i\bra{0}_i - \omega_i^{-}\ket{-1}_i\bra{-1}_i \\
 &+ \omega_i^{+}\ket{+1}_i\bra{+1}_i \\
\end{split} \\
& & \hspace{1.1em}H_{\rm ext} =\nu \hat{a}^\dag \hat{a} \label{ext}\\
& & H_{\rm couple} =\sum\limits_{i=1,2}\nu\eta_i(\hat{a}^\dag+\hat{a})\hat{\sigma}_{zi}
\label{gradient}
\end{eqnarray}
where all the Hamiltonians presented here are normalized by $\hbar$, $\omega_i^0$ and $\omega_i^\pm$ are the energies of the states $\ket{0}$ and $\ket{\pm 1}$ with respect to $\ket{0'}$, $\nu=\sqrt{3}\nu_z$ is the axial stretch mode frequency, $\hat{a}^\dag$ and $\hat{a}$ the creation and annihilation operators for that mode, $\hat{\sigma}_{zi}=\ket{+1}_i\bra{+1}_i-\ket{-1}_i\bra{-1}_i$, and $\eta_1$ and $\eta_2$ are the effective Lamb-Dicke parameters for the two ions, describing the strength of the coupling between the atoms' internal states and the mode due to the field gradient. $\eta_1 = -\eta_2 = z_0\mu_B\partial_zB/\sqrt{2}\hbar\nu$ where $\partial_zB$ is the axial magnetic-field gradient (the same for both ions) and $z_0=\sqrt{\hbar/2m\nu}$. For our system $\nu=2\pi\times459.34(1)$ kHz and $\partial_zB=23.6\,{\rm Tm}^{-1}$ giving $\eta_1=0.0041$. The couplings between \ket{0} and \ket{0'} and the motion, a consequence of the second-order Zeeman effect, are small and therefore are not considered.

\subsection*{Preparation and detection errors}

The magnetic-field gradient separates the frequencies $\omega_i^0$ of the clock transitions $\ket{0} \leftrightarrow \ket{0'}$ in the two ions by 11.9 kHz due to the second-order Zeeman shift. This allows the states $\ket{00}$, $\ket{00'}$, $\ket{0'0}$ and $\ket{0'0'}$ to be individually prepared using both optical pumping to prepare $\ket{00}$, and microwave $\pi$ pulses, resonant with the desired clock transition, to prepare $\ket{0'}$ \cite{Randall2}. We estimate that each of the states is prepared with infidelity less than $10^{-3}$. We measure the ion fluorescence using a photomultiplier tube and discriminate between the cases of 0, 1 and 2 ions fluorescing by setting two thresholds. We record the histograms after preparing each of the four spin states, allowing us to extract a linear map between the probabilities $P_0$, $P_1$ and $P_2$ obtained by thresholding, and the spin state probabilities $P_{00}$, $P_{00'}+P_{0'0}$ and $P_{0'0'}$. This mapping is then used to normalize data in subsequent experiments.

\subsection*{Motional coupling due to magnetic-field gradient}

he effect of the magnetic-field gradient is to allow transitions between internal states to affect the motional state of the ions. A microwave field oscillating at frequency $\omega_{\rm \mu w}=\omega^0_1+\omega^+_1\pm\nu+\delta$ close to one of the motional sidebands of the transition $\ket{0}_1\leftrightarrow\ket{+1}_1$, adds the following term to the Hamiltonian: $H_{\rm \mu w}=\Omuw(\ket{+1}_1\bra{0}_1+\ket{0}_1\bra{+1}_1)\cos{[\omega_{\rm \mu w}t]}$. By making a polaron-like (Schrieffer-Wolff) transformation \cite{polaron}

\begin{equation} 
U_p=\exp \bla{\sum_{i=1,2}\eta_i\blb{\hat{a}^\dagger - \hat{a}}\hat{\sigma}_{zi}},
\label{polaron}
\end{equation} the internal and external states are now coupled \cite{Mintert1, Gatis}, 
such that the driven term $H_{\rm \mu w}$ transforms to an (anti-) Jaynes-Cummings Hamiltonian

\begin{widetext}
\begin{equation}
H_{\rm \mu w} = \left\{\begin{array}{rcl}
\frac{\eta_1\Omuw}{2}(\ket{+1}_1\bra{0}_1\hat{a}^\dag e^{-i\delta t}+\ket{0}_1\bra{+1}_1\hat{a}e^{i\delta t}) & \text{for} &  \omega_{\rm \mu w}=\omega^0_1+\omega^+_1+\nu+\delta \\
-\frac{\eta_1\Omuw}{2}(\ket{+1}_1\bra{0}_1\hat{a} e^{-i\delta t}+\ket{0}_1\bra{+1}_1\hat{a}^\dag e^{i\delta t}) & \text{for} & \omega_{\rm \mu w}=\omega^0_1+\omega^+_1-\nu+\delta
\end{array}\right.
\end{equation}
\end{widetext}

\noindent depending on the field being tuned close to either the blue or red sideband, and where we have gone into the interaction picture and dropped all fast rotating terms.


\subsection*{Sideband cooling}
The experimental two-qubit gate sequence is preceded by a sideband cooling sequence similar to that described in Ref. \cite{Weidt}, however, here we use a microwave field instead of an RF field, to drive the red sideband. The ions are initially Doppler laser cooled for 4 ms using near-resonant light at 369 nm and prepared in $\ket{00}$ after 30 $\mu$s of optical pumping. The sideband cooling sequence then consists of applying a microwave field pulse of frequency $\omega_{\rm \mu w}=\omega^0_1+\omega^+_1-\nu$, driving the red sideband transition with a carrier Rabi frequency $\Omega/2\pi=74\,$kHz. Optical re-pumping then reinitialises the ions in $\ket{00}$. We apply a total of 500 repetitions of this sideband cooling sequence, each repetition applying an increasing microwave sideband pulse time which corresponds to the sideband Rabi frequencies of different populated $n$ levels. Using this sequence we achieve a final temperature of $\bar{n}=0.14(3)$.

\subsection*{Tunable quantum engineered clock qubit}

Qubits formed of the states $\ket{+1}_i$, $\ket{0}_i$ would rapidly decohere due to magnetic field fluctuations. Countering the magnetic field noise is done by applying four microwave fields of frequencies $\omega^0_1+\omega^+_1$, $\omega^0_1-\omega^-_1$, $\omega^0_2+\omega^+_2$, and $\omega^0_2-\omega^-_1$, resonant with the $\ket{+1}_1 \leftrightarrow \ket{0}_1$, $\ket{-1}_1 \leftrightarrow \ket{0}_1$, $\ket{+1}_2 \leftrightarrow \ket{0}_2$, and  $\ket{-1}_2 \leftrightarrow \ket{0}_2$ transitions respectively. If all four microwave fields are driven with equal Rabi frequencies $\Omuw$, the internal state Hamiltonian in the interaction picture becomes    
\begin{equation}\begin{split}
H_{\rm int} =&\frac{ \Omega_{\rm \mu w}}{2}\sum_{i=1,2}(\ket{0}_i\bra{+1}_i+\ket{0}_i\bra{-1}_i \\
&+ \ket{+1}_i\bra{0}_i+\ket{-1}_i\bra{0}_i),
\label{int}
\end{split}\end{equation}
and then transforms to the dressed state basis
\begin{equation}
H_{\rm int}=
\frac{\Omuw}{\sqrt{2}}\sum\limits_{i=1,2}(\ket{u}_i\bra{u}_i-\ket{d}_i\bra{d}_i),
\label{dressed}
\end{equation}
where $\ket{u}_i=\frac{1}{2}\ket{+1}_i+\frac{1}{2}\ket{-1}_i+\frac{1}{\sqrt{2}}\ket{0}_i$, $\ket{d}_i=\frac{1}{2}\ket{+1}_i+\frac{1}{2}\ket{-1}_i-\frac{1}{\sqrt{2}}\ket{0}_i$.  These resonantly driven transitions operate as a form of continuous dynamical decoupling for the $\Lambda$-system, and result in dark states $\ket{D}_i=\frac{1}{\sqrt{2}}(\ket{+1}_i-\ket{-1}_i)$, which are protected both against magnetic field noise and microwave amplitude fluctuations, such that together with the $\ket{0'}_i$ states constitute robust effective clock qubits \cite{Timoney,nati}.

Each qubit can be manipulated by an RF field 
\begin{equation}
H_{\rm rf}=\Omrf\bla{\ket{+1}_i\bra{0'}_i+\ket{0'}_i\bra{+1}_i}\cos\omega^{\rm rf}_i t.
\label{rf_0}
\end{equation}
 By setting $\omega^{\rm rf}_i=\omega^+_i$, and if $\Omrf\ll|\omega^+_i-\omega^-_i|$, then

\begin{equation}\label{Hrf2}\begin{split}
H_{\rm rf} & =  \frac{\Omrf}{2\sqrt{2}}\bla{\ket{D}_i\bra{0'}_i+\ket{0'}_i\bra{D}_i} \\
 & =  \frac{\Omega_0}{2}\bla{\ket{D}_i\bra{0'}_i+\ket{0'}_i\bra{D}_i},
\end{split}\end{equation}

\noindent in the interaction picture and after dropping fast rotating terms \cite{Webster}. We have defined a Rabi frequency $\Omega_0=\Omrf/\sqrt{2}$ which is the Rabi frequency for driving the engineered qubit.

The concept for a trapped-ion quantum computer presented in this manuscript requires well-protected qubits with a tunable transition frequency in the MHz range to allow for high-fidelity individual addressing within a set of global radiation fields. Traditionally, qubits consisting of a first-order magnetic field sensitive transition would have to be used, however such qubits are highly sensitive to ambient magnetic field fluctuations, limiting the achievable coherence time. As shown above the effective clock qubit which is well-protected from ambient magnetic field fluctuations can be manipulated by setting $\omega^{\rm rf}_j\approx\omega^+_i$ for ion $i$ and field $j$. Therefore, in order to change the transition frequency one simply changes the magnetic field environment at the ion to shift the first-order magnetic field sensitive state $\ket{+1}_i$ such that $\omega^+_i$ is resonant with the desired global radiation field.

\subsection*{Multi-qubit gate}
If instead of setting the RF to the resonant frequency $\omega^{\rm rf}_i=\omega^+_i$, four RF fields are set to be equally detuned from the red and blue sidebands, $\omega^{\rm rf}_i=\omega^+_i\pm(\nu+\delta)$, Eq. \ref{rf_0} in the interaction picture becomes
\begin{equation}
H_{\rm rf}=\Omega_{\rm rf}\sum_{i=1,2}\bla{\ket{+1}_i\bra{0'}_i+\ket{0'}_i\bra{+1}_i}\cos \blb{\bla{\nu+\delta}t}.
\label{rf1}
\end{equation}
This gives a M\o lmer-S\o rensen interaction, if coupling to the motional degrees of freedom is present. To show such coupling exists, a polaron-like transformation (Eq. \ref{polaron}) is made to the Hamiltonian containing the stretch mode (Eq. \ref{ext}), the magnetic-field gradient (Eq. \ref{gradient}), the microwave (Eq. \ref{int}) and the RF driving fields (Eq. \ref{rf1}),
\begin{equation}\begin{split}
H & = \nu \hat{a}^\dagger \hat{a} \\
 & +  \sum_{i=1,2}\nu \eta_i \bla{\hat{a}^\dagger + \hat{a} } \hat\sigma_{zi}\\
 &+ \frac{ \Omega_{\rm \mu w}}{2}\sum_{i=1,2}\bla{\ket{0}_i\bra{+1}_i+\ket{0}_i\bra{-1}_i +  \ket{+1}_i\bra{0}_i+\ket{-1}_i\bra{0}_i} \\
 &+ \Omega_{\rm rf} \sum_{i=1,2}\bla{\ket{+1}_i\bra{0'}_i+\ket{0'}_i\bra{+1}_i}\cos \blb{\bla{\nu+\delta}t},
\end{split}\end{equation}
which is therefore transformed to

\begin{equation}\label{eq:H0}\begin{split}
H_p&=U_p H U_p^\dagger =\nu \hat{a}^\dagger \hat{a}\\ 
&+\nu\sum_{i,j=1,2}\eta_i\eta_j\hat{\sigma}_{zi}\hat{\sigma}_{zj} \\ 
&+ \frac{\Omega_{\rm \mu w}}{2}\sum_{i=1,2}\blb{\bla{\ket{+1}_i\bra{0}_i+\ket{0}_i\bra{-1}_i} e^{\eta_i \bla{\hat{a}^\dagger - \hat{a} }}+{\rm h.c.}}\\
&+ \Omega_{\rm rf}\sum_{i=1,2}\bla{\ket{+1}_i\bra{0'}_ie^{\eta_i \bla{\hat{a}^\dagger - \hat{a} }}+{\rm h.c.}}\cos \blb{\bla{\nu+\delta}t}
\end{split}\end{equation}
where 
the RF fields in the last term couple the internal degrees of freedom to the external ones. In the Lamb-Dicke regime, where $\eta_i\sqrt{\bar{n}+1} \ll 1$,  the displacement operator $D(\eta_i)=e^{\eta_i\bla{\hat{a}^\dagger -\hat{a}}}$ is expanded in orders of the Lamb-Dicke parameter $\eta_i$. After transforming to the dressed state basis, and moving to the interaction picture with respect to the dressed state energy gap (Eq. \ref{dressed}) and the stretch mode (Eq. \ref{ext}), the first-order expansion of the RF transition term in Eq. \ref{eq:H0} yields
\begin{equation}\begin{split}
\Omega_{\rm rf}\sum_{i=1,2}&\eta_i\Bigg(\blb{\frac{\ket{u}_ie^{i\frac{\Omega_{\rm \mu w}}{\sqrt{2}}t} +\ket{d}_ie^{-i\frac{\Omega_{\rm \mu w}}{\sqrt{2}}t}  }{2} -\frac{\ket{D}_i}{\sqrt{2}}} \bra{0'}{_i} \\
&\times \bla{\hat{a}^\dagger e^{i\nu t}- \hat{a} e^{-i\nu t} }+{\rm h.c.}\Bigg)\cos \blb{\bla{\nu+\delta}t}.
\label{rf first}
\end{split}\end{equation}
Under the assumption $\delta \ll \frac{\Omega_{\rm \mu w}}{\sqrt{2}}\ll \nu$, the main contribution of Eq. \ref{rf first} is
\begin{equation}\begin{split}\label{Hgate}
H_{\rm gate} = &-\frac{\eta_1\Omega_0}{2}(\ket{D}_1\bra{0'}_1-\ket{0'}_1\bra{D}_1 \\
&-\ket{D}_2\bra{0'}_2 +\ket{0'}_2\bra{D}_2)\bla{\hat{a} e^{i\delta t}-\hat{a}^\dag e^{-i\delta t}},
\end{split}\end{equation}
which drives a M{\o}lmer-S{\o}rensen gate \cite{Sorensen2, Sorensen3}, whereas all the other terms are dropped in the rotating wave approximation. If such a Hamiltonian is applied for a time $\tau=2\pi/\delta$ then the qubit states are subject to the unitary transformation
\begin{equation}
U=\exp\blb{i\frac{\pi\eta_1^2\Omega_0^2}{\delta^2}\hat{\sigma}_{y1}\hat{\sigma}_{y2}},
\label{gate}
\end{equation}
with $\hat{\sigma}_{yi}=-i\bla{\ket{D}_i\bra{0'}_i-\ket{0'}_i\bra{D}_i}.$ By setting the detuning $\delta=2\eta_1\Omega_0$, the initial state $\ket{0'0'}$ is ideally transformed into the required maximally entangled state $\frac{1}{\sqrt{2}}(\ket{0'0'}-i\ket{DD})$.

\subsection*{Corrections to the gate}

The above derivation of $H_{\rm gate}$ considers only the slowest rotating terms. Terms dropped from the derivation lead to both sources of infidelity and  lightshifts of the qubit levels. We discuss these in this section, as well as methods to counteract such terms.

\subsubsection*{Off-resonant carrier excitation due to the rf driving fields}

In the standard M{\o}lmer-S{\o}rensen gate performed using trapped ions, off-resonant excitation of the carrier transitions can lead to a reduction in the average gate fidelity. 
There are a total of 6 RF-driven carrier transitions per ion, $\ket{0'}_i\leftrightarrow\ket{D}_i$, $\ket{0'}_i\leftrightarrow\ket{u}_i$ and $\ket{0'}_i\leftrightarrow\ket{d}_i$, each via both $\ket{+1}$ and $\ket{-1}$. The RF driven transitions originating from $\ket{0'}_i\leftrightarrow\ket{-1}$,  not considered in the derivation above due to the additional detuning of the second-order Zeeman splitting $\Delta_i=\omega^+_i-\omega^-_i \sim O(10\,{\rm kHz})$, result in a term
\begin{equation}\begin{split}
H_{\rm off}= &\frac{\Omega_{0}}{\sqrt{2}} \sum_{i=1,2} {\ket{-1}_i\bra{0'}_ie^{\eta_i \bla{a^\dagger e^{i\nu t} - a e^{-i\nu t} }}} \\
&\times \blb{e^{i\bla{\nu+\delta+\Delta_i}t}+e^{-i\bla{\nu+\delta-\Delta_i}t} }+{\rm h.c.}
\label{off}
\end{split}\end{equation}

\hfill \break 
As is the case with many other gate implementations, the effect of off-resonant carrier excitation is to produce a rapid oscillation in the gate fidelity.  Each carrier transition introduces an average infidelity of approximately $\Omega^2/\nu^2$, where $\Omega$ is the carrier Rabi frequency, which varies depending on if the transition is to \ket{D} or to \ket{u}/\ket{d} \cite{Webster}. Summing over the six carrier transitions, the total infidelity due to carrier excitation is $4\Omega_0^2/\nu^2$. In principle, it should be possible to time the gate to minimise the infidelity, however, this is difficult in practise. The standard approach to counteracting this infidelity of using pulse shaping of the gate fields \cite{Roos2} will work far more effectively for this gate, and is therefore implemented to effectively remove this effect. Instead of switching the gate pulses on and off near-instantaneously we shape the pulses in a way that at the beginning and end of the pulse the amplitude rises and falls adiabatically within a window length of $10 \us$ which was found to be sufficient using numerical simulations of the full gate dynamics.

\subsubsection*{Lightshifts and leakage outside the qubit subspace}
The RF-driven off-resonant carrier transitions in Eq. \ref{off} will produce a net lightshift term arising from non-vanishing A.C Stark shifts in the rotating frame of the dressed state energy gap in Eq. \ref{dressed}:
\begin{widetext}
\begin{equation}\begin{split}
H_{\rm rf_{-1}}^{\rm shift}=&\frac{\Omega_{0}^2}{4}\sum_{i=1,2}\bla{{\frac{1}{\nu+\Delta_i}}-{\frac{1}{\nu-\Delta_i}}}
\bla{\ket{D}_i\bra{D}_i-\ket{0'}_i\bra{0'}_i}\\
&-\frac{\Omega_{0}^2}{8}\sum_{i=1,2} \bigg(\frac{1}{\nu+\Omega_{\rm \mu w}/\sqrt{2}+\Delta_i}-\frac{1}{\nu-\Omega_{\rm \mu w}/\sqrt{2}-\Delta_i}
+ \frac{1}{\nu-\Omega_{\rm \mu w}/\sqrt{2}+\Delta_i}-\frac{1}{\nu+\Omega_{\rm \mu w}/\sqrt{2}-\Delta_i} \bigg)
\ket{0'}_i\bra{0'}_i,
\end{split}\end{equation}
\end{widetext} 
\hfill \break

\noindent where we have omitted $\delta$, since $\delta \ll\Delta_i , \Omega_{\rm \mu w}/\sqrt{2}, \nu $. This can be approximated as
\begin{equation}
 H_{\rm rf_{-1}}^{\rm shift} \approx
-\frac{3 \Omega_{0}^2}{4\nu^2}\sum_{i=1,2}\Delta_i\hat\sigma_{zi},
\end{equation}
under the assumption $\Delta_i , \Omega_{\rm \mu w}/\sqrt{2} \ll \nu $. The leading terms of the first-order expansion of the RF transitions in Eq. \ref{rf first} yield in the second order of perturbation the following terms:
\begin{equation}
H_{\rm rf}^{\rm shift}=\underbrace{\frac{\bla{\eta_1\Omega_0}^2}{2}\bla{\frac{1}{{\frac{\Omuw}{\sqrt{2}}-\delta}}+\frac{1}{{\frac{\Omuw}{\sqrt{2}}+\delta}}}}_{g_{\rm rf\ shift}} \sum_{i= 1,2}\ket{0'}_i\bra{0'}_i,
\label{rf-1_shift}\\
\end{equation}
which is a net lightshift term, as well as 
\begin{equation}\begin{split}
H_{\rm rf}^{\rm leak}&=\underbrace{\frac{\bla{\eta_1\Omega_0}^2}{2}\bla{\frac{1}{{\frac{\Omuw}{\sqrt{2}}-\delta}}-\frac{1}{{\frac{\Omuw}{\sqrt{2}}+\delta}}}}_{g_{\rm rf\ leak }} \\
&\times \bla{\ket{u}_1\bra{0'}_1-\ket{0'}_1\bra{d}_1}
\bla{\ket{0'}_2\bra{u}_2-\ket{d}_2\bra{0'}_2}+ {\rm h.c} 
\label{rf leakage}\\
\end{split}\end{equation}
which operates as a leakage outside of the qubit space. 

The magnetic-field gradient (second term in Eq. \ref{eq:H0}) together with the microwave driving fields (third term in Eq. \ref{eq:H0}) also produce similar contributions. After transforming to the dressed state basis and moving to the interaction picture with respect to the dressed state energy gap (Eq. \ref{dressed}) and the stretch mode (Eq. \ref{ext}), we obtain in second-order expansion of $\eta$
\begin{equation}\label{g+mw}\begin{split}
H_I&= -\sum_{i,j=1}^2 \frac{\nu\eta_i\eta_j}{4}\bla{\hat S_{+i}e^{i\frac{\Omega_{\rm \mu w}}{\sqrt{2}}t} +{\rm h.c.}}\bla{\hat S_{+j}e^{i\frac{\Omega_{\rm \mu w}}{\sqrt{2}}t} +{\rm h.c.}}\\
&-\frac{\Omega_{\rm \mu w}}{2\sqrt{2}}\sum_{i=1,2}\blb{\eta_i \bla{ \hat S_{+i}e^{i\frac{\Omega_{\rm \mu w}}{\sqrt{2}}t} -{\rm h.c.}} \bla{\hat{a}^\dagger e^{i\nu t} - \hat{a} e^{-i\nu t} } } 
\end{split}\end{equation}
where $\hat S_{+i}=\sqrt{2}(\ket{u}_i\bra{D}_i+\ket{D}_i\bra{d}_i)$ and $\hat S_{-i}=\hat S_{+i}^\dagger$. The leading unwanted terms originating from Eq. \ref{g+mw} give rise to
\begin{equation}
H_{\rm \mu w}^{\rm leak}=\underbrace{-\frac{\eta_1^2\nu^3}{2\bla{\frac{\Omuw^2}{2}-\nu^2}} }_{g_{\rm \mu w}} \bla{\hat S_{+1}\hat S_{-2}  +{\rm h.c}} , 
\label{gradient leakage}\\
\end{equation}
which leads to leakage outside of the qubit space, in addition to
\begin{equation}
H_{\rm \mu w}^{\rm shift}=\underbrace{\frac{\eta_1^2\nu^3}{\frac{\Omuw^2}{2}-\nu^2}}_{-2g_{\rm \mu w}}\bla{\ket{D}_1\bra{D}_1+\ket{D}_2\bra{D}_2},
\end{equation}
which is a lightshift term.

For the gate described in this manuscript, the coupling parameters for the leakage terms would be $g_{\rm rf\ leak}=2\pi\times0.05\,{\rm Hz}$ and $g_{\rm \mu w}=2\pi\times3.8\,{\rm Hz}$ and we find that $g_{\rm rf\ shift}=2\pi\times2.2\,{\rm Hz}$. We suppress their effect by introducing an additional energy gap between the coupled states, suppressing the couplings by making them off-resonant. We accomplish this by making the Rabi frequencies of the dressing fields for the two ions different by a small amount $\delta_0=\Omega_{{\rm \mu w}1} - \Omega_{{\rm \mu w}2}$ \cite{Gatis}. If $\delta_0\gg\eta^2\nu$, this coupling is energetically suppressed. The dressing field Rabi frequencies used for the gate presented here are $\Omega_{{\rm \mu w}1}=2\pi\times20.5\,$kHz and $\Omega_{{\rm \mu w}2}=2\pi\times21.6\,$kHz. The lightshifts are compensated by shifting the frequencies of the RF gate fields.

Expansion to higher orders in $\eta$ of Eq. \ref{off} yields the following terms 
\hfill \break
\begin{widetext}
\begin{equation}\begin{split}
H^{\rm higher}_{\rm rf}&=\frac{\Omega_{0}}{2\sqrt{2}}\sum_{j=1,2}\left(\eta_j\ket{u}_j\bra{0'}_j\bla{\hat{a}^\dagger e^{i\bla{\frac{\Omega_{\rm \mu w}}{\sqrt{2}}+\Delta_j-\delta}t}-\hat{a} e^{i\bla{\frac{\Omega_{\rm \mu w}}{\sqrt{2}}+\Delta_j+\delta}t}} +{\rm h.c.}\right)\\
&+\frac{\Omega_{0}}{2\sqrt{2}}\sum_{j=1,2}\left(\eta_j\ket{d}_j\bra{0'}_j\bla{\hat{a}^\dagger e^{i\bla{-\frac{\Omega_{\rm \mu w}}{\sqrt{2}}+\Delta_j-\delta}t}-\hat{a} e^{i\bla{-\frac{\Omega_{\rm \mu w}}{\sqrt{2}}+\Delta_j+\delta}t}} +{\rm h.c.}\right).
\end{split}\end{equation}\end{widetext}
These terms give rise to another A.C. Stark shift, also leading also to an additional lightshift term that is phonon-number dependent, and so cannot be compensated by a simple change in gate field frequencies, given by
\begin{equation}\begin{split}
H_{\rm phonon}^{\rm ls} = &\sum_{i=1,2}\underbrace{\frac{(\eta_1\Omega_0)^2}{8}\bla{\frac{1}{\frac{\Omega_{\rm \mu w}}{\sqrt{2}}-\Delta_i} -\frac{1}{\frac{\Omega_{\rm \mu w}}{\sqrt{2}}+\Delta_i}}}_{g_{\rm ph}}\\
\times &\bla{ 2\hat{a}^\dagger \hat{a} +1} \ket{0'}_i\bra{0'}_i.
\end{split}\end{equation}
The lightshift fluctuates from shot-to-shot due to the thermal spread in the phonon number, which scales as the square root of the mean phonon number. The strength of this term for our gate parameters is $g_{\rm ph}=2\pi\times 0.7\,$Hz and $2\pi\times 0.3\,$Hz for ions 1 and 2 respectively. The effect of this shift is already small, and could be further reduced by an increase in microwave Rabi frequency and reduction of the mode temperature.

\subsubsection*{Different Zeeman splitting due to the magnetic-field gradient}

The two ions are aligned along the magnetic-field gradient which determines the $z$-axis. Therefore they feel a different magnetic field, resulting in a difference in Zeeman splitting $\Delta_B=g\mu_B\partial_zB_z\Delta Z$, where $\Delta Z=\left(e^2/2\pi\epsilon_0M\nu^2 \right)^{1/3}$ is the distance between the two ions. This difference yields additional terms in the Hamiltonian in the rotating frame corresponding to the bare energy structure. Due to the four microwave driving fields the following additional terms are obtained:
\begin{equation}\begin{split}
H^{\rm \mu w}_{\rm Zeeman}=&\frac{\Omega_{\rm \mu w}}{2}( \left\vert +1\right\rangle_1 \left\langle 0\right\vert_1 +  \left\vert 0\right\rangle_1 \left\langle -1\right\vert_1 \\
&+ \left\vert -1\right\rangle_2 \left\langle 0\right\vert_2 +  \left\vert 0\right\rangle_2 \left\langle +1\right\vert_2  )e^{i\Delta_B t} +{\rm h.c.}
\label{Mw}
\end{split}\end{equation}
which in second order results in 
\begin{equation}
H^{\rm \mu w\ eff}_{\rm Zeeman}=\frac{\Omega_{\rm \mu w}^2}{4\Delta_B}\bla{F_{z1}-F_{z2}},
\label{mw devi}
\end{equation}
where $F_{zi}=\ket{+1}_i\bra{+1}_i-\ket{-1}_i\bra{-1}_i$. This should be taken into account when determining the microwave and RF frequencies.

Due to the four RF driving fields, the following terms should be added to the Hamiltonian:
\begin{equation}\label{RF}\begin{split}
H^{\rm rf}_{\rm Zeeman}&=\frac{\Omega_{0}}{\sqrt2}\bigg(\left(\ket{+1}_1 \bra{0'}_1+ \ket{0'}_2 \bra{+1}_2  \right) \\
&\times\bla{e^{i(\Delta_B+\nu+\delta) t}+e^{i(\Delta_B-\nu-\delta) t}} \\
&+\ket{-1}_1 \bra{0'}_1 \bla{e^{-i(\Delta_B-\Delta+\nu+\delta) t}+e^{-i(\Delta_B-\Delta-\nu-\delta) t} }\\
&+\ket{-1}_2 \bra{0'}_2 \bla{e^{i(\Delta_B+\Delta+\nu+\delta) t}+e^{i(\Delta_B+\Delta-\nu-\delta) t} }\\
&+{\rm h.c.}\bigg).
\end{split}\end{equation}
As a result, the four RF driving fields give rise to additional lightshifts given by
\begin{equation}
H^{\rm rf\ eff}_{\rm Zeeman}= \frac{\Delta \Omega_{0}^2}{\Delta_B^2-\nu^2} \bla{\ket{0'}_1\bra{0'}_1+\ket{0'}_2\bra{0'}_2}.
\label{rf deph}
\end{equation}
Together with the lightshift terms that were derived above, the net lightshift is compensated by shifting the frequencies of the gate fields.

\subsubsection*{Higher-order contribution to the gate transition}
An additional term arises when considering the following two terms in higher orders of perturbation: \\
(1.) the first-order expansion in $\eta$ of the microwave transition in Eq. \ref{eq:H0}
\begin{equation}
-\frac{\Omega_{\rm \mu w}}{2\sqrt{2}}\sum_{i=1,2}\eta_i\bla{ S_{+i} e^{i\frac{\Omega_{\rm \mu w}}{\sqrt{2}}t} -{\rm h.c.}  }\bla{ \hat{a}^\dagger e^{i\nu t} - {\rm h.c.}  }
\end{equation} 
(2.) the RF carrier transition (last term in Eq. \ref{eq:H0})
\begin{equation}\begin{split}
\frac{\Omega_{0}}{\sqrt2}\sum_{i=1,2}&\bla{\blb{\ket{u}_i \bra{0'}_i e^{i\frac{\Omega_{\rm \mu w}}{\sqrt{2}}t} + \ket{d}_i \bra{0'}_ie^{-i\frac{\Omega_{\rm \mu w}}{\sqrt{2}}t} } 
+{\rm h.c.}}\\
&\times\cos \blb{\bla{\nu+\delta}t}.
\end{split}\end{equation}
These two terms oscillate almost with the same frequency $\approx \nu\pm \Omega_{\rm \mu w}/\sqrt{2}$, where the difference is exactly $\delta$, such that Raman transitions are obtained
\begin{equation}\begin{split}
H_{\rm h.o.}&=\frac{\Omega_{\rm \mu w}^2}{\Omega_{\rm \mu w}^2-2\nu^2} \sum_{j=1,2}\frac{\eta_j\Omega_{0}}{2} \bla{\ket{D}_j \bra{0'}_j - \ket{0'}_j \bra{D}_j   } \\
& \hspace{10em}\times\bla{ \hat{a} e^{i\delta t} - \hat{a}^\dagger e^{-i\delta t} }\\
&=\frac{\Omega_{\rm \mu w}^2}{2\nu^2-\Omega_{\rm \mu w}^2} H_{\rm gate}
\end{split}\end{equation}
This produces a M{\o}lmer-S{\o}rensen coupling equivalent to Eq. \ref{Hgate}, that changes the required gate duration as the Rabi frequency of the microwave driving fields $\Omega_{\rm \mu w}$ is increased. In the limit $\Omega_{\rm \mu w}  \gg \nu$, this term yields an opposite sign to the desired coupling, cancelling the gate completely due to destructive interference with these Raman transitions. In our experiment $\Omega_{\rm \mu w}  \ll \nu$, and the change in gate time is negligible.

\subsubsection*{Imperfections in the dressing fields}

Introducing a small imbalance in the amplitudes of the two microwave driving fields $\Delta\Omega_{\rm \mu w_i}=\Omega_{\rm \mu w_i}^+-\Omega_{\rm \mu w_i}^-$, with $\Delta\Omega_{\rm \mu w_i} \ll \Omega_{\rm \mu w_i}$,  yields an additional term:
\begin{equation}
\sum_{i=1,2}\frac{\Delta\Omega_{\rm \mu w_i}}{2}\bla{\ket{+1}_i\bra{0}_i - \ket{-1}_i\bra{0}_i+ {\rm h.c.}  }.
\end{equation}
In the interaction picture with respect to the dressed state energy, this yields an A.C. Stark shift which does not operate in our qubit subspace:
\begin{equation}
\sum_{i=1,2}\frac{\Delta\Omega_{\rm \mu w_i}^2}{2\sqrt{2}\Omega_{\rm \mu w_i}}\bla{\ket{u}_i\bra{u}_i - \ket{d}_i\bra{d}_i}.
\end{equation}
However, this amplitude imbalance together with the ambient magnetic field fluctuations $\delta B(t)$ gives rise to another noise-inducing term that survives the RWA: 
\begin{equation}\begin{split}
\sum_{i=1,2}&\frac{\sqrt{2}\bla{\frac{\Delta\Omega_{\rm \mu w_i}} {2}-\frac{\delta B}{\sqrt{2}}  }^2 }{\Omega_{\rm \mu w_i}}\ket{u}_i\bra{u}_i \\
&-\frac{\sqrt{2}\bla{\frac{\Delta\Omega_{\rm \mu w_i}} {2}+\frac{\delta B}{\sqrt{2}}  }^2 }{\Omega_{\rm \mu w_i}}\ket{d}_i\bra{d}_i \\
& + \frac{2\delta B\Delta\Omega_{\rm \mu w_i}}{\Omega_{\rm \mu w_i}}\ket{D}_i\bra{D}_i.
\end{split}\end{equation}
The first two terms do not operate in our qubit subspace, while the latter is another $\sigma_{zi}$ lightshift term coupled to the ambient magnetic noise, and therefore causes dephasing.  This term equals exactly the original dephasing term multiplied by $ \frac{\Delta\Omega_{\rm \mu w_i}}{\Omega_{\rm \mu w_i}}$, which means that the original dephasing term is being prolonged by a factor of $\left( \frac{\Delta\Omega_{\rm \mu w_i}}{\Omega_{\rm \mu w_i}} \right)^2$. This factor is typically smaller than $10^{-4}$. 

\subsection*{Further increasing the gate fidelity}

The infidelity of our demonstrated gate is dominated by heating of the motional mode, and depolarization of the qubit. Increasing the speed of the gate reduces both sources of infidelity. In addition, the heating rate can be further reduced by increasing the trap frequency. By modifying our gate parameters (including the size of the magnetic field gradient) the infidelities can be dramatically reduced and fault-tolerant operation is easily achievable.

We have modeled the following parameters: $\Omega_{\rm \mu w}/2 \pi =10\,$kHz , $\Omega_{\rm 0}/2 \pi =198\,$kHz, $\nu/2 \pi =1.1\,$MHz, $\eta =0.0071$ (which corresponds to a magnetic field gradient $\partial_z B_z = 150\,$T/m), with pulse shaping using a $\sin^2(t)$ profile with rise and fall duration $t_{shaping}=10\pi/\nu$. Instead of applying an imbalance between $\Omega_{\rm \mu w1}$ and $\Omega_{\rm \mu w2}$, we have detuned the microwave transitions equally by $2\pi\times 0.5\,$kHz with respect to the $\ket{0}_i$ levels \cite{Gatis}. To check these changes do not substantially increase the gate's intrinsic infidelities discussed above, we have simulated the gate performance, namely the gate state fidelity for each one of the four possible states in the code space $\blc{ \ket{D,D},\ket{D,0'},\ket{0',D},\ket{0',0'}}$, from which we calculate the gate process fidelity as their average. In the simulation, we consider a vibrational mode with a cutoff $n_{cut}=15$ and no further approximations have been made. Taking into account a depolarisation time of 2 s (as previously measured using our dressed-state system \cite{Baumgart}) as well as a stretch-mode heating rate of 1.3$\,{\rm s}^{-1}$ (heating rate as measured in the relevant apparatus and scaled to 1.1 MHz; if using an ion chip with 150 $\mu$m ion-electrode distance, the expected tenfold increase in heating rate could be compensated by light cooling of the trap electrodes to liquid nitrogen temperature \cite{Deslauriers1}), we calculate a total fidelity of 0.999 and a gate time of 361$\mu s$. Further improvement is of course possible with longer coherence times and lower heating rates of the type seen in other experiments.

\subsection*{Extension to a large-scale architecture}

Our method can be applied to construct a large scale quantum computer. We have developed a detailed engineering blueprint for this purpose \cite{Lekitsch}. Here we discuss important considerations relevant to our method. Individual addressing of ions in the same entanglement zone is achieved using the local magnetic field gradients, while individual addressing of ions in different zones is achieved by applying local voltages to position ions in different zones in a different magnetic offset field. The crosstalk between ions $i$ and $j$ for a square pulse resonant with ion $i$ can be characterised by the time-averaged excitation probability of ion $j$, given by $C_{ij} \approx \Omega_j^2/2\Delta_{ij}^2$, where $\Omega_j$ is the Rabi frequency of the desired transition in ion $j$ and $\Delta_{ij}$ is the frequency separation between the transitions in the two ions \cite{Piltz}. In a single entanglement zone, the frequency separation between the Zeeman sublevels of the ion pair for the parameters in the example case in the previous section is 9.8 MHz. For the microwave dressing fields, Rabi frequencies $\Omega_{\rm \mu w}/2\pi = 10$ kHz are used, and therefore the crosstalk values for the dressing fields are $C_{12} = C_{21} = 5.2 \times 10^{-7}$. For the rf fields used to drive the two-qubit gate, the crosstalk values would be $C_{12} = C_{21} = 2.0 \times 10^{-4}$ for square pulse shapes, however, the fields amplitudes would be shaped with a $\sin^2$ profile as demonstrated in this work. Shaping the pulse amplitudes further reduces the crosstalk by several orders of magnitude. To see this, a numerical simulation of a two-level system driven by a field with Rabi frequency $\Omega(t)$ and detuning $\delta$ was performed. The Rabi frequency was varied in time starting with a $\sin^2$ shape ramp from $\Omega = 0$ to $\Omega = \Omega_\text{max}$ for a time $t_\text{w}$, followed by a hold at $\Omega_\text{max}$ for a time $t_\text{h}$, and finally a second $\sin^2$ shape ramp down to $\Omega = 0$ in time $t_\text{w}$. It was found that for  $\delta > 10 \Omega_\text{max}$ and $t_\text{w} > \pi/ \Omega_\text{max}$, the error is reduced to $< 10^{-7}$. This detuning requirement is fulfilled for both the rf and microwave dressing fields in this example. Therefore the crosstalk between ions in a single zone is $< 10^{-6}$, and is therefore negligible compared to other error sources. 

As mentioned, individual addressing of ions in different zones is achieved by positioning the ions in different zones in a different local static magnetic offset field achieved making use of the position dependent magnetic field originating from the local static magnetic field gradient within each zone. Ions that are not being addressed sit at magnetic field $B_1$ corresponding to position $z_1$, while ions that require to be addressed are moved to position $z_2$ resulting in a magnetic field $B_2$. As an example, if $B_2-B_1 = 2$ G, the Zeeman states of the ions that are not being addressed are 2.8 MHz off-resonant. The crosstalk for such a frequency separation with the parameters in the example case is $6.4 \times 10^{-6}$ for the microwave dressing fields, and $< 10^{-7}$ for the shaped rf gate field pulse. Other types of gates can then be introduced by positioning the ions in different locations resulting in additional magnetic offset fields $B_3$, $B_4$ etc. The minimal set of gates required for a universal quantum computer following the surface code error correction scheme described in Ref. \cite{Fowler} consists of two single qubit gates (Hadamard + $\pi/8$ $\sigma_z$-rotation) and a two-qubit entangling gate such as the one presented in this work. Therefore there are four offset magnetic fields required: No interaction, single qubit Hadamard, single qubit $\pi/8$ $\sigma_z$-rotation and two-qubit gate. The total required range of magnetic field offsets is therefore approximately 6 G for an arbitrarily large processor. Additional operations could be added by increasing the range of magnetic field offsets if required.

The currents creating the static magnetic field gradients local to each gate zone are applied permanently and are not switched on or off. An alternative method to select arbitrary gate zones for gate execution is to add an additional current carrying wire to each gate zone. The low current passing through this wire in each gate zone creating the required different levels of magnetic field need to be switched in order to individually address ions in different zones. Here, currents of $\approx 100$ mA need to be applied to the wires to ramp the local offset B-field from $B_1$ to $B_2$ (a difference of $\approx 2$ G) which would then shift the qubit frequency into resonance with a particular set of global gate fields. Different levels of current are applied to relevant coils to shift the qubit frequencies into resonance with different global gate fields. On-chip digital-to-analogue converters can be used to control the currents with 2 MS/s and 16 bit precision. A realistic timing sequence for a two-qubit gate operation for example would then be a 5 $\mu s$ ramp from $B_1$ to $B_2$, followed by the gate operation which is then followed by a second 5 $\mu s$ ramp from $B_2$ back to $B_1$ where an integrated filter produces a smooth waveform.

\bibliography{microwave_gate}

\begin{thebibliography}{48}%
\makeatletter
\providecommand \@ifxundefined [1]{%
 \@ifx{#1\undefined}
}%
\providecommand \@ifnum [1]{%
 \ifnum #1\expandafter \@firstoftwo
 \else \expandafter \@secondoftwo
 \fi
}%
\providecommand \@ifx [1]{%
 \ifx #1\expandafter \@firstoftwo
 \else \expandafter \@secondoftwo
 \fi
}%
\providecommand \natexlab [1]{#1}%
\providecommand \enquote  [1]{``#1''}%
\providecommand \bibnamefont  [1]{#1}%
\providecommand \bibfnamefont [1]{#1}%
\providecommand \citenamefont [1]{#1}%
\providecommand \href@noop [0]{\@secondoftwo}%
\providecommand \href [0]{\begingroup \@sanitize@url \@href}%
\providecommand \@href[1]{\@@startlink{#1}\@@href}%
\providecommand \@@href[1]{\endgroup#1\@@endlink}%
\providecommand \@sanitize@url [0]{\catcode `\\12\catcode `\$12\catcode
  `\&12\catcode `\#12\catcode `\^12\catcode `\_12\catcode `\%12\relax}%
\providecommand \@@startlink[1]{}%
\providecommand \@@endlink[0]{}%
\providecommand \url  [0]{\begingroup\@sanitize@url \@url }%
\providecommand \@url [1]{\endgroup\@href {#1}{\urlprefix }}%
\providecommand \urlprefix  [0]{URL }%
\providecommand \Eprint [0]{\href }%
\providecommand \doibase [0]{http://dx.doi.org/}%
\providecommand \selectlanguage [0]{\@gobble}%
\providecommand \bibinfo  [0]{\@secondoftwo}%
\providecommand \bibfield  [0]{\@secondoftwo}%
\providecommand \translation [1]{[#1]}%
\providecommand \BibitemOpen [0]{}%
\providecommand \bibitemStop [0]{}%
\providecommand \bibitemNoStop [0]{.\EOS\space}%
\providecommand \EOS [0]{\spacefactor3000\relax}%
\providecommand \BibitemShut  [1]{\csname bibitem#1\endcsname}%
\let\auto@bib@innerbib\@empty
\bibitem [{\citenamefont {Blatt}\ and\ \citenamefont
  {Wineland}(2008)}]{Blatt1}%
  \BibitemOpen
  \bibfield  {author} {\bibinfo {author} {\bibfnamefont {R.}~\bibnamefont
  {Blatt}}\ and\ \bibinfo {author} {\bibfnamefont {D.}~\bibnamefont
  {Wineland}},\ }\href {\doibase 10.1038/nature07125} {\bibfield  {journal}
  {\bibinfo  {journal} {Nature}\ }\textbf {\bibinfo {volume} {453}},\ \bibinfo
  {pages} {1008} (\bibinfo {year} {2008})}\BibitemShut {NoStop}%
\bibitem [{\citenamefont {H\"{a}ffner}\ \emph {et~al.}(2005)\citenamefont
  {H\"{a}ffner}, \citenamefont {H\"{a}nsel}, \citenamefont {Roos},
  \citenamefont {Benhelm}, \citenamefont {al~kar}, \citenamefont {Chwalla},
  \citenamefont {K\"{o}rber}, \citenamefont {Rapol}, \citenamefont {Riebe},
  \citenamefont {Schmidt}, \citenamefont {Becher}, \citenamefont {G\"{u}hne},
  \citenamefont {D\"{u}r},\ and\ \citenamefont {Blatt}}]{Haffner3}%
  \BibitemOpen
  \bibfield  {author} {\bibinfo {author} {\bibfnamefont {H.}~\bibnamefont
  {H\"{a}ffner}}, \bibinfo {author} {\bibfnamefont {W.}~\bibnamefont
  {H\"{a}nsel}}, \bibinfo {author} {\bibfnamefont {C.~F.}\ \bibnamefont
  {Roos}}, \bibinfo {author} {\bibfnamefont {J.}~\bibnamefont {Benhelm}},
  \bibinfo {author} {\bibfnamefont {D.~C.}\ \bibnamefont {al~kar}}, \bibinfo
  {author} {\bibfnamefont {M.}~\bibnamefont {Chwalla}}, \bibinfo {author}
  {\bibfnamefont {T.}~\bibnamefont {K\"{o}rber}}, \bibinfo {author}
  {\bibfnamefont {U.~D.}\ \bibnamefont {Rapol}}, \bibinfo {author}
  {\bibfnamefont {M.}~\bibnamefont {Riebe}}, \bibinfo {author} {\bibfnamefont
  {P.~O.}\ \bibnamefont {Schmidt}}, \bibinfo {author} {\bibfnamefont
  {C.}~\bibnamefont {Becher}}, \bibinfo {author} {\bibfnamefont
  {O.}~\bibnamefont {G\"{u}hne}}, \bibinfo {author} {\bibfnamefont
  {W.}~\bibnamefont {D\"{u}r}}, \ and\ \bibinfo {author} {\bibfnamefont
  {R.}~\bibnamefont {Blatt}},\ }\href@noop {} {\bibfield  {journal} {\bibinfo
  {journal} {Nature}\ }\textbf {\bibinfo {volume} {438}},\ \bibinfo {pages}
  {643} (\bibinfo {year} {2005})}\BibitemShut {NoStop}%
\bibitem [{\citenamefont {Sacket}\ \emph {et~al.}(2000)\citenamefont {Sacket},
  \citenamefont {Kielpinski}, \citenamefont {King}, \citenamefont {Langer},
  \citenamefont {Meyer}, \citenamefont {Myatt}, \citenamefont {Rowe},
  \citenamefont {Turchette}, \citenamefont {Itano}, \citenamefont {Wineland},\
  and\ \citenamefont {Monroe}}]{Sackett}%
  \BibitemOpen
  \bibfield  {author} {\bibinfo {author} {\bibfnamefont {C.~A.}\ \bibnamefont
  {Sacket}}, \bibinfo {author} {\bibfnamefont {D.}~\bibnamefont {Kielpinski}},
  \bibinfo {author} {\bibfnamefont {B.~E.}\ \bibnamefont {King}}, \bibinfo
  {author} {\bibfnamefont {C.}~\bibnamefont {Langer}}, \bibinfo {author}
  {\bibfnamefont {V.}~\bibnamefont {Meyer}}, \bibinfo {author} {\bibfnamefont
  {C.~J.}\ \bibnamefont {Myatt}}, \bibinfo {author} {\bibfnamefont
  {M.}~\bibnamefont {Rowe}}, \bibinfo {author} {\bibfnamefont {Q.~A.}\
  \bibnamefont {Turchette}}, \bibinfo {author} {\bibfnamefont {W.~M.}\
  \bibnamefont {Itano}}, \bibinfo {author} {\bibfnamefont {D.~J.}\ \bibnamefont
  {Wineland}}, \ and\ \bibinfo {author} {\bibfnamefont {C.}~\bibnamefont
  {Monroe}},\ }\href@noop {} {\bibfield  {journal} {\bibinfo  {journal}
  {Nature}\ }\textbf {\bibinfo {volume} {404}},\ \bibinfo {pages} {256}
  (\bibinfo {year} {2000})}\BibitemShut {NoStop}%
\bibitem [{\citenamefont {Leibfried}\ \emph {et~al.}(2005)\citenamefont
  {Leibfried}, \citenamefont {Knill}, \citenamefont {Seidlin}, \citenamefont
  {Britton}, \citenamefont {Blakestad}, \citenamefont {Chiaverini},
  \citenamefont {Hume}, \citenamefont {Itano}, \citenamefont {Jost},
  \citenamefont {Langer}, \citenamefont {Ozeri}, \citenamefont {Reichle},\ and\
  \citenamefont {Wineland}}]{Leibfried3}%
  \BibitemOpen
  \bibfield  {author} {\bibinfo {author} {\bibfnamefont {D.}~\bibnamefont
  {Leibfried}}, \bibinfo {author} {\bibfnamefont {E.}~\bibnamefont {Knill}},
  \bibinfo {author} {\bibfnamefont {S.}~\bibnamefont {Seidlin}}, \bibinfo
  {author} {\bibfnamefont {J.}~\bibnamefont {Britton}}, \bibinfo {author}
  {\bibfnamefont {R.~B.}\ \bibnamefont {Blakestad}}, \bibinfo {author}
  {\bibfnamefont {J.}~\bibnamefont {Chiaverini}}, \bibinfo {author}
  {\bibfnamefont {D.~B.}\ \bibnamefont {Hume}}, \bibinfo {author}
  {\bibfnamefont {W.~M.}\ \bibnamefont {Itano}}, \bibinfo {author}
  {\bibfnamefont {J.~D.}\ \bibnamefont {Jost}}, \bibinfo {author}
  {\bibfnamefont {C.}~\bibnamefont {Langer}}, \bibinfo {author} {\bibfnamefont
  {R.}~\bibnamefont {Ozeri}}, \bibinfo {author} {\bibfnamefont
  {R.}~\bibnamefont {Reichle}}, \ and\ \bibinfo {author} {\bibfnamefont
  {D.~J.}\ \bibnamefont {Wineland}},\ }\href@noop {} {\bibfield  {journal}
  {\bibinfo  {journal} {Nature}\ }\textbf {\bibinfo {volume} {438}},\ \bibinfo
  {pages} {639} (\bibinfo {year} {2005})}\BibitemShut {NoStop}%
\bibitem [{\citenamefont {Blatt}\ and\ \citenamefont {Roos}(2012)}]{Blatt}%
  \BibitemOpen
  \bibfield  {author} {\bibinfo {author} {\bibfnamefont {R.}~\bibnamefont
  {Blatt}}\ and\ \bibinfo {author} {\bibfnamefont {C.~F.}\ \bibnamefont
  {Roos}},\ }\href@noop {} {\bibfield  {journal} {\bibinfo  {journal} {Nat
  Phys}\ }\textbf {\bibinfo {volume} {8}},\ \bibinfo {pages} {277} (\bibinfo
  {year} {2012})}\BibitemShut {NoStop}%
\bibitem [{\citenamefont {Friedenauer}\ \emph {et~al.}(2008)\citenamefont
  {Friedenauer}, \citenamefont {Schmitz}, \citenamefont {Glueckert},
  \citenamefont {Porras},\ and\ \citenamefont {Schaetz}}]{Friedenauer}%
  \BibitemOpen
  \bibfield  {author} {\bibinfo {author} {\bibfnamefont {A.}~\bibnamefont
  {Friedenauer}}, \bibinfo {author} {\bibfnamefont {H.}~\bibnamefont
  {Schmitz}}, \bibinfo {author} {\bibfnamefont {J.~T.}\ \bibnamefont
  {Glueckert}}, \bibinfo {author} {\bibfnamefont {D.}~\bibnamefont {Porras}}, \
  and\ \bibinfo {author} {\bibfnamefont {T.}~\bibnamefont {Schaetz}},\
  }\href@noop {} {\bibfield  {journal} {\bibinfo  {journal} {Nature Phys.}\
  }\textbf {\bibinfo {volume} {4}},\ \bibinfo {pages} {757} (\bibinfo {year}
  {2008})}\BibitemShut {NoStop}%
\bibitem [{\citenamefont {Kim}\ \emph {et~al.}(2010)\citenamefont {Kim},
  \citenamefont {Chang}, \citenamefont {Korenblit}, \citenamefont {Islam},
  \citenamefont {Edwards}, \citenamefont {Freericks}, \citenamefont {Lin},
  \citenamefont {Duan},\ and\ \citenamefont {Monroe}}]{Kim}%
  \BibitemOpen
  \bibfield  {author} {\bibinfo {author} {\bibfnamefont {K.}~\bibnamefont
  {Kim}}, \bibinfo {author} {\bibfnamefont {M.-S.}\ \bibnamefont {Chang}},
  \bibinfo {author} {\bibfnamefont {S.}~\bibnamefont {Korenblit}}, \bibinfo
  {author} {\bibfnamefont {R.}~\bibnamefont {Islam}}, \bibinfo {author}
  {\bibfnamefont {E.~E.}\ \bibnamefont {Edwards}}, \bibinfo {author}
  {\bibfnamefont {J.~K.}\ \bibnamefont {Freericks}}, \bibinfo {author}
  {\bibfnamefont {G.-D.}\ \bibnamefont {Lin}}, \bibinfo {author} {\bibfnamefont
  {L.-M.}\ \bibnamefont {Duan}}, \ and\ \bibinfo {author} {\bibfnamefont
  {C.}~\bibnamefont {Monroe}},\ }\href@noop {} {\bibfield  {journal} {\bibinfo
  {journal} {Nature}\ }\textbf {\bibinfo {volume} {465}},\ \bibinfo {pages}
  {590} (\bibinfo {year} {2010})}\BibitemShut {NoStop}%
\bibitem [{\citenamefont {Lanyon}\ \emph {et~al.}(2011)\citenamefont {Lanyon}
  \emph {et~al.}}]{Lanyon}%
  \BibitemOpen
  \bibfield  {author} {\bibinfo {author} {\bibfnamefont {B.~P.}\ \bibnamefont
  {Lanyon}} \emph {et~al.},\ }\href {\doibase 10.1126/science.1208001}
  {\bibfield  {journal} {\bibinfo  {journal} {Science}\ }\textbf {\bibinfo
  {volume} {334}},\ \bibinfo {pages} {57} (\bibinfo {year} {2011})}\BibitemShut
  {NoStop}%
\bibitem [{\citenamefont {Johanning}\ \emph {et~al.}(2009)\citenamefont
  {Johanning}, \citenamefont {Var\'on},\ and\ \citenamefont
  {Wunderlich}}]{Johanning}%
  \BibitemOpen
  \bibfield  {author} {\bibinfo {author} {\bibfnamefont {M.}~\bibnamefont
  {Johanning}}, \bibinfo {author} {\bibfnamefont {A.~F.}\ \bibnamefont
  {Var\'on}}, \ and\ \bibinfo {author} {\bibfnamefont {C.}~\bibnamefont
  {Wunderlich}},\ }\href@noop {} {\bibfield  {journal} {\bibinfo  {journal}
  {Journal of Physics B: Atomic, Molecular and Optical Physics}\ }\textbf
  {\bibinfo {volume} {42}},\ \bibinfo {pages} {154009} (\bibinfo {year}
  {2009})}\BibitemShut {NoStop}%
\bibitem [{\citenamefont {Schneider}\ \emph {et~al.}(2012)\citenamefont
  {Schneider}, \citenamefont {Porras},\ and\ \citenamefont
  {Schaetz}}]{Schneider1}%
  \BibitemOpen
  \bibfield  {author} {\bibinfo {author} {\bibfnamefont {C.}~\bibnamefont
  {Schneider}}, \bibinfo {author} {\bibfnamefont {D.}~\bibnamefont {Porras}}, \
  and\ \bibinfo {author} {\bibfnamefont {T.}~\bibnamefont {Schaetz}},\
  }\href@noop {} {\bibfield  {journal} {\bibinfo  {journal} {Reports on
  Progress in Physics}\ }\textbf {\bibinfo {volume} {75}},\ \bibinfo {pages}
  {024401} (\bibinfo {year} {2012})}\BibitemShut {NoStop}%
\bibitem [{\citenamefont {Ludlow}\ \emph {et~al.}(2015)\citenamefont {Ludlow},
  \citenamefont {Boyd}, \citenamefont {Ye}, \citenamefont {Peik},\ and\
  \citenamefont {Schmidt}}]{Ludlow}%
  \BibitemOpen
  \bibfield  {author} {\bibinfo {author} {\bibfnamefont {A.~D.}\ \bibnamefont
  {Ludlow}}, \bibinfo {author} {\bibfnamefont {M.~M.}\ \bibnamefont {Boyd}},
  \bibinfo {author} {\bibfnamefont {J.}~\bibnamefont {Ye}}, \bibinfo {author}
  {\bibfnamefont {E.}~\bibnamefont {Peik}}, \ and\ \bibinfo {author}
  {\bibfnamefont {P.~O.}\ \bibnamefont {Schmidt}},\ }\href {\doibase
  10.1103/RevModPhys.87.637} {\bibfield  {journal} {\bibinfo  {journal} {Rev.
  Mod. Phys.}\ }\textbf {\bibinfo {volume} {87}},\ \bibinfo {pages} {637}
  (\bibinfo {year} {2015})}\BibitemShut {NoStop}%
\bibitem [{\citenamefont {Kotler}\ \emph {et~al.}(2011)\citenamefont {Kotler},
  \citenamefont {Akerman}, \citenamefont {Glickman}, \citenamefont {Keselman},\
  and\ \citenamefont {Ozeri}}]{Kotler}%
  \BibitemOpen
  \bibfield  {author} {\bibinfo {author} {\bibfnamefont {S.}~\bibnamefont
  {Kotler}}, \bibinfo {author} {\bibfnamefont {N.}~\bibnamefont {Akerman}},
  \bibinfo {author} {\bibfnamefont {Y.}~\bibnamefont {Glickman}}, \bibinfo
  {author} {\bibfnamefont {A.}~\bibnamefont {Keselman}}, \ and\ \bibinfo
  {author} {\bibfnamefont {R.}~\bibnamefont {Ozeri}},\ }\href {\doibase
  doi:10.1038/nature10010} {\bibfield  {journal} {\bibinfo  {journal} {Nature}\
  }\textbf {\bibinfo {volume} {473}},\ \bibinfo {pages} {61} (\bibinfo {year}
  {2011})}\BibitemShut {NoStop}%
\bibitem [{\citenamefont {Pruttivarasin}\ \emph {et~al.}(2015)\citenamefont
  {Pruttivarasin}, \citenamefont {Ramm}, \citenamefont {Porsev}, \citenamefont
  {Tupitsyn}, \citenamefont {Safronova}, \citenamefont {Hohensee},\ and\
  \citenamefont {H\"affner}}]{Pruttivarasin}%
  \BibitemOpen
  \bibfield  {author} {\bibinfo {author} {\bibfnamefont {T.}~\bibnamefont
  {Pruttivarasin}}, \bibinfo {author} {\bibfnamefont {M.}~\bibnamefont {Ramm}},
  \bibinfo {author} {\bibfnamefont {S.~G.}\ \bibnamefont {Porsev}}, \bibinfo
  {author} {\bibfnamefont {I.~I.}\ \bibnamefont {Tupitsyn}}, \bibinfo {author}
  {\bibfnamefont {M.~S.}\ \bibnamefont {Safronova}}, \bibinfo {author}
  {\bibfnamefont {M.~A.}\ \bibnamefont {Hohensee}}, \ and\ \bibinfo {author}
  {\bibfnamefont {H.}~\bibnamefont {H\"affner}},\ }\href {\doibase
  doi:10.1038/nature14091} {\bibfield  {journal} {\bibinfo  {journal} {Nature}\
  }\textbf {\bibinfo {volume} {517}},\ \bibinfo {pages} {592} (\bibinfo {year}
  {2015})}\BibitemShut {NoStop}%
\bibitem [{\citenamefont {Baumgart}\ \emph {et~al.}(2014)\citenamefont
  {Baumgart}, \citenamefont {Cai}, \citenamefont {Retzker}, \citenamefont
  {Plenio},\ and\ \citenamefont {Wunderlich}}]{Baumgart}%
  \BibitemOpen
  \bibfield  {author} {\bibinfo {author} {\bibfnamefont {I.}~\bibnamefont
  {Baumgart}}, \bibinfo {author} {\bibfnamefont {J.~M.}\ \bibnamefont {Cai}},
  \bibinfo {author} {\bibfnamefont {A.}~\bibnamefont {Retzker}}, \bibinfo
  {author} {\bibfnamefont {M.~B.}\ \bibnamefont {Plenio}}, \ and\ \bibinfo
  {author} {\bibfnamefont {C.}~\bibnamefont {Wunderlich}},\ }\href
  {http://arxiv.org/abs/1411.7893} {\bibfield  {journal} {\bibinfo  {journal}
  {arXiv:1411.7893}\ } (\bibinfo {year} {2014})}\BibitemShut {NoStop}%
\bibitem [{\citenamefont {H\"affner}\ \emph {et~al.}(2008)\citenamefont
  {H\"affner}, \citenamefont {Roos},\ and\ \citenamefont {Blatt}}]{Haffner2}%
  \BibitemOpen
  \bibfield  {author} {\bibinfo {author} {\bibfnamefont {H.}~\bibnamefont
  {H\"affner}}, \bibinfo {author} {\bibfnamefont {C.~F.}\ \bibnamefont {Roos}},
  \ and\ \bibinfo {author} {\bibfnamefont {R.}~\bibnamefont {Blatt}},\
  }\href@noop {} {\bibfield  {journal} {\bibinfo  {journal} {Physical Reports}\
  }\textbf {\bibinfo {volume} {469}},\ \bibinfo {pages} {155} (\bibinfo {year}
  {2008})}\BibitemShut {NoStop}%
\bibitem [{\citenamefont {Leibfried}\ \emph {et~al.}(2003)\citenamefont
  {Leibfried}, \citenamefont {DeMarco}, \citenamefont {Meyer}, \citenamefont
  {Lucas}, \citenamefont {Barrett}, \citenamefont {Britton}, \citenamefont
  {Itano}, \citenamefont {Jelenkovic}, \citenamefont {Langer}, \citenamefont
  {Rosenband},\ and\ \citenamefont {Wineland}}]{Leibfried2}%
  \BibitemOpen
  \bibfield  {author} {\bibinfo {author} {\bibfnamefont {D.}~\bibnamefont
  {Leibfried}}, \bibinfo {author} {\bibfnamefont {B.}~\bibnamefont {DeMarco}},
  \bibinfo {author} {\bibfnamefont {V.}~\bibnamefont {Meyer}}, \bibinfo
  {author} {\bibfnamefont {D.}~\bibnamefont {Lucas}}, \bibinfo {author}
  {\bibfnamefont {M.}~\bibnamefont {Barrett}}, \bibinfo {author} {\bibfnamefont
  {J.}~\bibnamefont {Britton}}, \bibinfo {author} {\bibfnamefont {W.~M.}\
  \bibnamefont {Itano}}, \bibinfo {author} {\bibfnamefont {B.}~\bibnamefont
  {Jelenkovic}}, \bibinfo {author} {\bibfnamefont {C.}~\bibnamefont {Langer}},
  \bibinfo {author} {\bibfnamefont {T.}~\bibnamefont {Rosenband}}, \ and\
  \bibinfo {author} {\bibfnamefont {D.~J.}\ \bibnamefont {Wineland}},\ }\href
  {\doibase 10.1038/nature01492} {\bibfield  {journal} {\bibinfo  {journal}
  {Nature}\ }\textbf {\bibinfo {volume} {422}},\ \bibinfo {pages} {412}
  (\bibinfo {year} {2003})}\BibitemShut {NoStop}%
\bibitem [{\citenamefont {Cirac}\ and\ \citenamefont {Zoller}(2000)}]{Cirac}%
  \BibitemOpen
  \bibfield  {author} {\bibinfo {author} {\bibfnamefont {J.~I.}\ \bibnamefont
  {Cirac}}\ and\ \bibinfo {author} {\bibfnamefont {P.}~\bibnamefont {Zoller}},\
  }\href@noop {} {\bibfield  {journal} {\bibinfo  {journal} {Nature}\ }\textbf
  {\bibinfo {volume} {404}},\ \bibinfo {pages} {579} (\bibinfo {year}
  {2000})}\BibitemShut {NoStop}%
\bibitem [{\citenamefont {Kielpinski}\ \emph {et~al.}(2002)\citenamefont
  {Kielpinski}, \citenamefont {Monroe},\ and\ \citenamefont
  {Wineland}}]{Kielpinski}%
  \BibitemOpen
  \bibfield  {author} {\bibinfo {author} {\bibfnamefont {D.}~\bibnamefont
  {Kielpinski}}, \bibinfo {author} {\bibfnamefont {C.}~\bibnamefont {Monroe}},
  \ and\ \bibinfo {author} {\bibfnamefont {D.}~\bibnamefont {Wineland}},\
  }\href@noop {} {\bibfield  {journal} {\bibinfo  {journal} {Nature}\ }\textbf
  {\bibinfo {volume} {417}},\ \bibinfo {pages} {709} (\bibinfo {year}
  {2002})}\BibitemShut {NoStop}%
\bibitem [{\citenamefont {Monroe}\ and\ \citenamefont {Kim}(2013)}]{Monroe4}%
  \BibitemOpen
  \bibfield  {author} {\bibinfo {author} {\bibfnamefont {C.}~\bibnamefont
  {Monroe}}\ and\ \bibinfo {author} {\bibfnamefont {J.}~\bibnamefont {Kim}},\
  }\href {\doibase 10.1126/science.1231298} {\bibfield  {journal} {\bibinfo
  {journal} {Science}\ }\textbf {\bibinfo {volume} {339}},\ \bibinfo {pages}
  {1164} (\bibinfo {year} {2013})}\BibitemShut {NoStop}%
\bibitem [{\citenamefont {Nielsen}\ and\ \citenamefont
  {Chuang}(2010)}]{Nielsen}%
  \BibitemOpen
  \bibfield  {author} {\bibinfo {author} {\bibfnamefont {M.~A.}\ \bibnamefont
  {Nielsen}}\ and\ \bibinfo {author} {\bibfnamefont {I.~L.}\ \bibnamefont
  {Chuang}},\ }\href@noop {} {\emph {\bibinfo {title} {Quantum computation and
  quantum information}}}\ (\bibinfo  {publisher} {Cambridge University Press},\
  \bibinfo {year} {2010})\BibitemShut {NoStop}%
\bibitem [{\citenamefont {Akerman}\ \emph {et~al.}(2015)\citenamefont
  {Akerman}, \citenamefont {Navon}, \citenamefont {Kotler}, \citenamefont
  {Glickman},\ and\ \citenamefont {Ozeri}}]{Akerman}%
  \BibitemOpen
  \bibfield  {author} {\bibinfo {author} {\bibfnamefont {N.}~\bibnamefont
  {Akerman}}, \bibinfo {author} {\bibfnamefont {N.}~\bibnamefont {Navon}},
  \bibinfo {author} {\bibfnamefont {S.}~\bibnamefont {Kotler}}, \bibinfo
  {author} {\bibfnamefont {Y.}~\bibnamefont {Glickman}}, \ and\ \bibinfo
  {author} {\bibfnamefont {R.}~\bibnamefont {Ozeri}},\ }\href@noop {}
  {\bibfield  {journal} {\bibinfo  {journal} {New Journal of Physics}\ }\textbf
  {\bibinfo {volume} {17}},\ \bibinfo {pages} {113060} (\bibinfo {year}
  {2015})}\BibitemShut {NoStop}%
\bibitem [{\citenamefont {Ballance}\ \emph {et~al.}(2015)\citenamefont
  {Ballance}, \citenamefont {Harty}, \citenamefont {Linke}, \citenamefont
  {Sepiol},\ and\ \citenamefont {Lucas}}]{Ballance}%
  \BibitemOpen
  \bibfield  {author} {\bibinfo {author} {\bibfnamefont {C.~J.}\ \bibnamefont
  {Ballance}}, \bibinfo {author} {\bibfnamefont {T.~P.}\ \bibnamefont {Harty}},
  \bibinfo {author} {\bibfnamefont {N.~M.}\ \bibnamefont {Linke}}, \bibinfo
  {author} {\bibfnamefont {M.~A.}\ \bibnamefont {Sepiol}}, \ and\ \bibinfo
  {author} {\bibfnamefont {D.~M.}\ \bibnamefont {Lucas}},\ }\href
  {http://arxiv.org/abs/1512.04600} {\bibfield  {journal} {\bibinfo  {journal}
  {arXiv:1512.04600}\ } (\bibinfo {year} {2015})}\BibitemShut {NoStop}%
\bibitem [{\citenamefont {Gaebler}\ \emph {et~al.}(2014)\citenamefont
  {Gaebler}, \citenamefont {Tan}, \citenamefont {Lin}, \citenamefont {Wan},
  \citenamefont {Bowler}, \citenamefont {Keith}, \citenamefont {Glancy},
  \citenamefont {Coakley}, \citenamefont {Knill}, \citenamefont {Leibfried},\
  and\ \citenamefont {Wineland}}]{Gaebler}%
  \BibitemOpen
  \bibfield  {author} {\bibinfo {author} {\bibfnamefont {J.~P.}\ \bibnamefont
  {Gaebler}}, \bibinfo {author} {\bibfnamefont {T.~R.}\ \bibnamefont {Tan}},
  \bibinfo {author} {\bibfnamefont {Y.}~\bibnamefont {Lin}}, \bibinfo {author}
  {\bibfnamefont {Y.}~\bibnamefont {Wan}}, \bibinfo {author} {\bibfnamefont
  {R.}~\bibnamefont {Bowler}}, \bibinfo {author} {\bibfnamefont {A.~C.}\
  \bibnamefont {Keith}}, \bibinfo {author} {\bibfnamefont {S.}~\bibnamefont
  {Glancy}}, \bibinfo {author} {\bibfnamefont {K.}~\bibnamefont {Coakley}},
  \bibinfo {author} {\bibfnamefont {E.}~\bibnamefont {Knill}}, \bibinfo
  {author} {\bibfnamefont {D.}~\bibnamefont {Leibfried}}, \ and\ \bibinfo
  {author} {\bibfnamefont {D.~J.}\ \bibnamefont {Wineland}},\ }\href
  {http://arxiv.org/abs/1604.00032} {\bibfield  {journal} {\bibinfo  {journal}
  {arXiv:1604.00032}\ } (\bibinfo {year} {2014})}\BibitemShut {NoStop}%
\bibitem [{\citenamefont {Harty}\ \emph {et~al.}(2014)\citenamefont {Harty},
  \citenamefont {Allcock}, \citenamefont {Ballance}, \citenamefont {Guidoni},
  \citenamefont {Janacek}, \citenamefont {Linke}, \citenamefont {Stacey},\ and\
  \citenamefont {Lucas}}]{Harty}%
  \BibitemOpen
  \bibfield  {author} {\bibinfo {author} {\bibfnamefont {T.~P.}\ \bibnamefont
  {Harty}}, \bibinfo {author} {\bibfnamefont {D.~T.~C.}\ \bibnamefont
  {Allcock}}, \bibinfo {author} {\bibfnamefont {C.~J.}\ \bibnamefont
  {Ballance}}, \bibinfo {author} {\bibfnamefont {L.}~\bibnamefont {Guidoni}},
  \bibinfo {author} {\bibfnamefont {H.~A.}\ \bibnamefont {Janacek}}, \bibinfo
  {author} {\bibfnamefont {N.~M.}\ \bibnamefont {Linke}}, \bibinfo {author}
  {\bibfnamefont {D.~N.}\ \bibnamefont {Stacey}}, \ and\ \bibinfo {author}
  {\bibfnamefont {D.~M.}\ \bibnamefont {Lucas}},\ }\href {\doibase
  10.1103/PhysRevLett.113.220501} {\bibfield  {journal} {\bibinfo  {journal}
  {Phys. Rev. Lett.}\ }\textbf {\bibinfo {volume} {113}},\ \bibinfo {pages}
  {220501} (\bibinfo {year} {2014})}\BibitemShut {NoStop}%
\bibitem [{\citenamefont {Mintert}\ and\ \citenamefont
  {Wunderlich}(2001)}]{Mintert1}%
  \BibitemOpen
  \bibfield  {author} {\bibinfo {author} {\bibfnamefont {F.}~\bibnamefont
  {Mintert}}\ and\ \bibinfo {author} {\bibfnamefont {C.}~\bibnamefont
  {Wunderlich}},\ }\href {\doibase 10.1103/PhysRevLett.87.257904} {\bibfield
  {journal} {\bibinfo  {journal} {Phys. Rev. Lett.}\ }\textbf {\bibinfo
  {volume} {87}},\ \bibinfo {pages} {257904} (\bibinfo {year}
  {2001})}\BibitemShut {NoStop}%
\bibitem [{\citenamefont {Ospelkaus}\ \emph {et~al.}(2008)\citenamefont
  {Ospelkaus}, \citenamefont {Langer}, \citenamefont {Amini}, \citenamefont
  {Brown}, \citenamefont {Leibfried},\ and\ \citenamefont
  {Wineland}}]{Ospelkaus1}%
  \BibitemOpen
  \bibfield  {author} {\bibinfo {author} {\bibfnamefont {C.}~\bibnamefont
  {Ospelkaus}}, \bibinfo {author} {\bibfnamefont {C.~E.}\ \bibnamefont
  {Langer}}, \bibinfo {author} {\bibfnamefont {J.~M.}\ \bibnamefont {Amini}},
  \bibinfo {author} {\bibfnamefont {K.~R.}\ \bibnamefont {Brown}}, \bibinfo
  {author} {\bibfnamefont {D.}~\bibnamefont {Leibfried}}, \ and\ \bibinfo
  {author} {\bibfnamefont {D.~J.}\ \bibnamefont {Wineland}},\ }\href {\doibase
  10.1103/PhysRevLett.101.090502} {\bibfield  {journal} {\bibinfo  {journal}
  {Phys. Rev. Lett.}\ }\textbf {\bibinfo {volume} {101}},\ \bibinfo {pages}
  {090502} (\bibinfo {year} {2008})}\BibitemShut {NoStop}%
\bibitem [{\citenamefont {Ospelkaus}\ \emph {et~al.}(2011)\citenamefont
  {Ospelkaus}, \citenamefont {Warring}, \citenamefont {Colombe}, \citenamefont
  {Brown}, \citenamefont {Amini}, \citenamefont {Leibfried},\ and\
  \citenamefont {Wineland}}]{Ospelkaus}%
  \BibitemOpen
  \bibfield  {author} {\bibinfo {author} {\bibfnamefont {C.}~\bibnamefont
  {Ospelkaus}}, \bibinfo {author} {\bibfnamefont {U.}~\bibnamefont {Warring}},
  \bibinfo {author} {\bibfnamefont {Y.}~\bibnamefont {Colombe}}, \bibinfo
  {author} {\bibfnamefont {K.~R.}\ \bibnamefont {Brown}}, \bibinfo {author}
  {\bibfnamefont {J.~M.}\ \bibnamefont {Amini}}, \bibinfo {author}
  {\bibfnamefont {D.}~\bibnamefont {Leibfried}}, \ and\ \bibinfo {author}
  {\bibfnamefont {D.~J.}\ \bibnamefont {Wineland}},\ }\href@noop {} {\bibfield
  {journal} {\bibinfo  {journal} {Nature}\ }\textbf {\bibinfo {volume} {476}},\
  \bibinfo {pages} {181} (\bibinfo {year} {2011})}\BibitemShut {NoStop}%
\bibitem [{\citenamefont {Warring}\ \emph {et~al.}(2013)\citenamefont
  {Warring}, \citenamefont {Ospelkaus}, \citenamefont {Colombe}, \citenamefont
  {J\"ordens}, \citenamefont {Leibfried},\ and\ \citenamefont
  {Wineland}}]{Warring}%
  \BibitemOpen
  \bibfield  {author} {\bibinfo {author} {\bibfnamefont {U.}~\bibnamefont
  {Warring}}, \bibinfo {author} {\bibfnamefont {C.}~\bibnamefont {Ospelkaus}},
  \bibinfo {author} {\bibfnamefont {Y.}~\bibnamefont {Colombe}}, \bibinfo
  {author} {\bibfnamefont {R.}~\bibnamefont {J\"ordens}}, \bibinfo {author}
  {\bibfnamefont {D.}~\bibnamefont {Leibfried}}, \ and\ \bibinfo {author}
  {\bibfnamefont {D.~J.}\ \bibnamefont {Wineland}},\ }\href {\doibase
  10.1103/PhysRevLett.110.173002} {\bibfield  {journal} {\bibinfo  {journal}
  {Phys. Rev. Lett.}\ }\textbf {\bibinfo {volume} {110}},\ \bibinfo {pages}
  {173002} (\bibinfo {year} {2013})}\BibitemShut {NoStop}%
\bibitem [{\citenamefont {Craik}\ \emph {et~al.}(2016)\citenamefont {Craik},
  \citenamefont {Linke}, \citenamefont {Sepiol}, \citenamefont {Harty},
  \citenamefont {Ballance}, \citenamefont {Stacey}, \citenamefont {Steane},
  \citenamefont {Lucas},\ and\ \citenamefont {Allcock}}]{Craik}%
  \BibitemOpen
  \bibfield  {author} {\bibinfo {author} {\bibfnamefont {D.~P. L.~A.}\
  \bibnamefont {Craik}}, \bibinfo {author} {\bibfnamefont {N.~M.}\ \bibnamefont
  {Linke}}, \bibinfo {author} {\bibfnamefont {M.~A.}\ \bibnamefont {Sepiol}},
  \bibinfo {author} {\bibfnamefont {T.~P.}\ \bibnamefont {Harty}}, \bibinfo
  {author} {\bibfnamefont {C.~J.}\ \bibnamefont {Ballance}}, \bibinfo {author}
  {\bibfnamefont {D.~N.}\ \bibnamefont {Stacey}}, \bibinfo {author}
  {\bibfnamefont {A.~M.}\ \bibnamefont {Steane}}, \bibinfo {author}
  {\bibfnamefont {D.~M.}\ \bibnamefont {Lucas}}, \ and\ \bibinfo {author}
  {\bibfnamefont {D.~T.~C.}\ \bibnamefont {Allcock}},\ }\href
  {http://arxiv.org/abs/1601.02696} {\bibfield  {journal} {\bibinfo  {journal}
  {arXiv:1601.02696}\ } (\bibinfo {year} {2016})}\BibitemShut {NoStop}%
\bibitem [{\citenamefont {Khromova}\ \emph {et~al.}(2012)\citenamefont
  {Khromova}, \citenamefont {Piltz}, \citenamefont {Scharfenberger},
  \citenamefont {Gloger}, \citenamefont {Johanning}, \citenamefont {Var\'on},\
  and\ \citenamefont {Wunderlich}}]{Khromova1}%
  \BibitemOpen
  \bibfield  {author} {\bibinfo {author} {\bibfnamefont {A.}~\bibnamefont
  {Khromova}}, \bibinfo {author} {\bibfnamefont {C.}~\bibnamefont {Piltz}},
  \bibinfo {author} {\bibfnamefont {B.}~\bibnamefont {Scharfenberger}},
  \bibinfo {author} {\bibfnamefont {T.~F.}\ \bibnamefont {Gloger}}, \bibinfo
  {author} {\bibfnamefont {M.}~\bibnamefont {Johanning}}, \bibinfo {author}
  {\bibfnamefont {A.~F.}\ \bibnamefont {Var\'on}}, \ and\ \bibinfo {author}
  {\bibfnamefont {C.}~\bibnamefont {Wunderlich}},\ }\href {\doibase
  10.1103/PhysRevLett.108.220502} {\bibfield  {journal} {\bibinfo  {journal}
  {Phys. Rev. Lett.}\ }\textbf {\bibinfo {volume} {108}},\ \bibinfo {pages}
  {220502} (\bibinfo {year} {2012})}\BibitemShut {NoStop}%
\bibitem [{\citenamefont {Timoney}\ \emph {et~al.}(2011)\citenamefont
  {Timoney}, \citenamefont {Baumgart}, \citenamefont {Johanning}, \citenamefont
  {Varon}, \citenamefont {Plenio}, \citenamefont {Retzker},\ and\ \citenamefont
  {Wunderlich}}]{Timoney}%
  \BibitemOpen
  \bibfield  {author} {\bibinfo {author} {\bibfnamefont {N.}~\bibnamefont
  {Timoney}}, \bibinfo {author} {\bibfnamefont {I.}~\bibnamefont {Baumgart}},
  \bibinfo {author} {\bibfnamefont {M.}~\bibnamefont {Johanning}}, \bibinfo
  {author} {\bibfnamefont {A.~F.}\ \bibnamefont {Varon}}, \bibinfo {author}
  {\bibfnamefont {M.~B.}\ \bibnamefont {Plenio}}, \bibinfo {author}
  {\bibfnamefont {A.}~\bibnamefont {Retzker}}, \ and\ \bibinfo {author}
  {\bibfnamefont {C.}~\bibnamefont {Wunderlich}},\ }\href {\doibase
  10.1038/nature10319} {\bibfield  {journal} {\bibinfo  {journal} {Nature}\
  }\textbf {\bibinfo {volume} {476}},\ \bibinfo {pages} {185} (\bibinfo {year}
  {2011})}\BibitemShut {NoStop}%
\bibitem [{\citenamefont {Tan}\ \emph {et~al.}(2013)\citenamefont {Tan},
  \citenamefont {Gaebler}, \citenamefont {Bowler}, \citenamefont {Lin},
  \citenamefont {Jost}, \citenamefont {Leibfried},\ and\ \citenamefont
  {Wineland}}]{Tan}%
  \BibitemOpen
  \bibfield  {author} {\bibinfo {author} {\bibfnamefont {T.~R.}\ \bibnamefont
  {Tan}}, \bibinfo {author} {\bibfnamefont {J.~P.}\ \bibnamefont {Gaebler}},
  \bibinfo {author} {\bibfnamefont {R.}~\bibnamefont {Bowler}}, \bibinfo
  {author} {\bibfnamefont {Y.}~\bibnamefont {Lin}}, \bibinfo {author}
  {\bibfnamefont {J.~D.}\ \bibnamefont {Jost}}, \bibinfo {author}
  {\bibfnamefont {D.}~\bibnamefont {Leibfried}}, \ and\ \bibinfo {author}
  {\bibfnamefont {D.~J.}\ \bibnamefont {Wineland}},\ }\href {\doibase
  10.1103/PhysRevLett.110.263002} {\bibfield  {journal} {\bibinfo  {journal}
  {Phys. Rev. Lett.}\ }\textbf {\bibinfo {volume} {110}},\ \bibinfo {pages}
  {263002} (\bibinfo {year} {2013})}\BibitemShut {NoStop}%
\bibitem [{\citenamefont {Webster}\ \emph {et~al.}(2013)\citenamefont
  {Webster}, \citenamefont {Weidt}, \citenamefont {Lake}, \citenamefont
  {McLoughlin},\ and\ \citenamefont {Hensinger}}]{Webster}%
  \BibitemOpen
  \bibfield  {author} {\bibinfo {author} {\bibfnamefont {S.~C.}\ \bibnamefont
  {Webster}}, \bibinfo {author} {\bibfnamefont {S.}~\bibnamefont {Weidt}},
  \bibinfo {author} {\bibfnamefont {K.}~\bibnamefont {Lake}}, \bibinfo {author}
  {\bibfnamefont {J.~J.}\ \bibnamefont {McLoughlin}}, \ and\ \bibinfo {author}
  {\bibfnamefont {W.~K.}\ \bibnamefont {Hensinger}},\ }\href {\doibase
  10.1103/PhysRevLett.111.140501} {\bibfield  {journal} {\bibinfo  {journal}
  {Phys. Rev. Lett.}\ }\textbf {\bibinfo {volume} {111}},\ \bibinfo {pages}
  {140501} (\bibinfo {year} {2013})}\BibitemShut {NoStop}%
\bibitem [{\citenamefont {Weidt}\ \emph {et~al.}(2015)\citenamefont {Weidt},
  \citenamefont {Randall}, \citenamefont {Webster}, \citenamefont {Standing},
  \citenamefont {Rodriguez}, \citenamefont {Webb}, \citenamefont {Lekitsch},\
  and\ \citenamefont {Hensinger}}]{Weidt}%
  \BibitemOpen
  \bibfield  {author} {\bibinfo {author} {\bibfnamefont {S.}~\bibnamefont
  {Weidt}}, \bibinfo {author} {\bibfnamefont {J.}~\bibnamefont {Randall}},
  \bibinfo {author} {\bibfnamefont {S.~C.}\ \bibnamefont {Webster}}, \bibinfo
  {author} {\bibfnamefont {E.~D.}\ \bibnamefont {Standing}}, \bibinfo {author}
  {\bibfnamefont {A.}~\bibnamefont {Rodriguez}}, \bibinfo {author}
  {\bibfnamefont {A.~E.}\ \bibnamefont {Webb}}, \bibinfo {author}
  {\bibfnamefont {B.}~\bibnamefont {Lekitsch}}, \ and\ \bibinfo {author}
  {\bibfnamefont {W.~K.}\ \bibnamefont {Hensinger}},\ }\href {\doibase
  10.1103/PhysRevLett.115.013002} {\bibfield  {journal} {\bibinfo  {journal}
  {Phys. Rev. Lett.}\ }\textbf {\bibinfo {volume} {115}},\ \bibinfo {pages}
  {013002} (\bibinfo {year} {2015})}\BibitemShut {NoStop}%
\bibitem [{\citenamefont {Lekitsch}\ \emph {et~al.}(2015)\citenamefont
  {Lekitsch}, \citenamefont {Weidt}, \citenamefont {Fowler}, \citenamefont
  {M\o{}lmer}, \citenamefont {Devitt}, \citenamefont {Wunderlich},\ and\
  \citenamefont {Hensinger}}]{Lekitsch}%
  \BibitemOpen
  \bibfield  {author} {\bibinfo {author} {\bibfnamefont {B.}~\bibnamefont
  {Lekitsch}}, \bibinfo {author} {\bibfnamefont {S.}~\bibnamefont {Weidt}},
  \bibinfo {author} {\bibfnamefont {A.~G.}\ \bibnamefont {Fowler}}, \bibinfo
  {author} {\bibfnamefont {K.}~\bibnamefont {M\o{}lmer}}, \bibinfo {author}
  {\bibfnamefont {S.~J.}\ \bibnamefont {Devitt}}, \bibinfo {author}
  {\bibfnamefont {C.}~\bibnamefont {Wunderlich}}, \ and\ \bibinfo {author}
  {\bibfnamefont {W.~K.}\ \bibnamefont {Hensinger}},\ }\href
  {http://arxiv.org/abs/1508.00420} {\bibfield  {journal} {\bibinfo  {journal}
  {arXiv:1508.00420}\ } (\bibinfo {year} {2015})}\BibitemShut {NoStop}%
\bibitem [{\citenamefont {Monroe}\ \emph {et~al.}(2014)\citenamefont {Monroe},
  \citenamefont {Raussendorf}, \citenamefont {Ruthven}, \citenamefont {Brown},
  \citenamefont {Maunz}, \citenamefont {Duan},\ and\ \citenamefont
  {Kim}}]{Monroe3}%
  \BibitemOpen
  \bibfield  {author} {\bibinfo {author} {\bibfnamefont {C.}~\bibnamefont
  {Monroe}}, \bibinfo {author} {\bibfnamefont {R.}~\bibnamefont {Raussendorf}},
  \bibinfo {author} {\bibfnamefont {A.}~\bibnamefont {Ruthven}}, \bibinfo
  {author} {\bibfnamefont {K.~R.}\ \bibnamefont {Brown}}, \bibinfo {author}
  {\bibfnamefont {P.}~\bibnamefont {Maunz}}, \bibinfo {author} {\bibfnamefont
  {L.-M.}\ \bibnamefont {Duan}}, \ and\ \bibinfo {author} {\bibfnamefont
  {J.}~\bibnamefont {Kim}},\ }\href {\doibase 10.1103/PhysRevA.89.022317}
  {\bibfield  {journal} {\bibinfo  {journal} {Phys. Rev. A}\ }\textbf {\bibinfo
  {volume} {89}},\ \bibinfo {pages} {022317} (\bibinfo {year}
  {2014})}\BibitemShut {NoStop}%
\bibitem [{\citenamefont {S\o{}rensen}\ and\ \citenamefont
  {M\o{}lmer}(1999)}]{Sorensen2}%
  \BibitemOpen
  \bibfield  {author} {\bibinfo {author} {\bibfnamefont {A.}~\bibnamefont
  {S\o{}rensen}}\ and\ \bibinfo {author} {\bibfnamefont {K.}~\bibnamefont
  {M\o{}lmer}},\ }\href {\doibase 10.1103/PhysRevLett.82.1971} {\bibfield
  {journal} {\bibinfo  {journal} {Phys. Rev. Lett.}\ }\textbf {\bibinfo
  {volume} {82}},\ \bibinfo {pages} {1971} (\bibinfo {year}
  {1999})}\BibitemShut {NoStop}%
\bibitem [{\citenamefont {McLoughlin}\ \emph {et~al.}(2011)\citenamefont
  {McLoughlin}, \citenamefont {Nizamani}, \citenamefont {Siverns},
  \citenamefont {Sterling}, \citenamefont {Hughes}, \citenamefont {Lekitsch},
  \citenamefont {Stein}, \citenamefont {Weidt},\ and\ \citenamefont
  {Hensinger}}]{McLoughlin2}%
  \BibitemOpen
  \bibfield  {author} {\bibinfo {author} {\bibfnamefont {J.~J.}\ \bibnamefont
  {McLoughlin}}, \bibinfo {author} {\bibfnamefont {A.~H.}\ \bibnamefont
  {Nizamani}}, \bibinfo {author} {\bibfnamefont {J.~D.}\ \bibnamefont
  {Siverns}}, \bibinfo {author} {\bibfnamefont {R.~C.}\ \bibnamefont
  {Sterling}}, \bibinfo {author} {\bibfnamefont {M.~D.}\ \bibnamefont
  {Hughes}}, \bibinfo {author} {\bibfnamefont {B.}~\bibnamefont {Lekitsch}},
  \bibinfo {author} {\bibfnamefont {B.}~\bibnamefont {Stein}}, \bibinfo
  {author} {\bibfnamefont {S.}~\bibnamefont {Weidt}}, \ and\ \bibinfo {author}
  {\bibfnamefont {W.~K.}\ \bibnamefont {Hensinger}},\ }\href {\doibase
  10.1103/PhysRevA.83.013406} {\bibfield  {journal} {\bibinfo  {journal} {Phys.
  Rev. A}\ }\textbf {\bibinfo {volume} {83}},\ \bibinfo {pages} {013406}
  (\bibinfo {year} {2011})}\BibitemShut {NoStop}%
\bibitem [{\citenamefont {Lake}\ \emph {et~al.}(2015)\citenamefont {Lake},
  \citenamefont {Weidt}, \citenamefont {Randall}, \citenamefont {Standing},
  \citenamefont {Webster},\ and\ \citenamefont {Hensinger}}]{Lake}%
  \BibitemOpen
  \bibfield  {author} {\bibinfo {author} {\bibfnamefont {K.}~\bibnamefont
  {Lake}}, \bibinfo {author} {\bibfnamefont {S.}~\bibnamefont {Weidt}},
  \bibinfo {author} {\bibfnamefont {J.}~\bibnamefont {Randall}}, \bibinfo
  {author} {\bibfnamefont {E.~D.}\ \bibnamefont {Standing}}, \bibinfo {author}
  {\bibfnamefont {S.~C.}\ \bibnamefont {Webster}}, \ and\ \bibinfo {author}
  {\bibfnamefont {W.~K.}\ \bibnamefont {Hensinger}},\ }\href {\doibase
  10.1103/PhysRevA.91.012319} {\bibfield  {journal} {\bibinfo  {journal} {Phys.
  Rev. A}\ }\textbf {\bibinfo {volume} {91}},\ \bibinfo {pages} {012319}
  (\bibinfo {year} {2015})}\BibitemShut {NoStop}%
\bibitem [{\citenamefont {Randall}\ \emph {et~al.}(2015)\citenamefont
  {Randall}, \citenamefont {Weidt}, \citenamefont {Standing}, \citenamefont
  {Lake}, \citenamefont {Webster}, \citenamefont {Murgia}, \citenamefont
  {Navickas}, \citenamefont {Roth},\ and\ \citenamefont
  {Hensinger}}]{Randall2}%
  \BibitemOpen
  \bibfield  {author} {\bibinfo {author} {\bibfnamefont {J.}~\bibnamefont
  {Randall}}, \bibinfo {author} {\bibfnamefont {S.}~\bibnamefont {Weidt}},
  \bibinfo {author} {\bibfnamefont {E.~D.}\ \bibnamefont {Standing}}, \bibinfo
  {author} {\bibfnamefont {K.}~\bibnamefont {Lake}}, \bibinfo {author}
  {\bibfnamefont {S.~C.}\ \bibnamefont {Webster}}, \bibinfo {author}
  {\bibfnamefont {D.~F.}\ \bibnamefont {Murgia}}, \bibinfo {author}
  {\bibfnamefont {T.}~\bibnamefont {Navickas}}, \bibinfo {author}
  {\bibfnamefont {K.}~\bibnamefont {Roth}}, \ and\ \bibinfo {author}
  {\bibfnamefont {W.~K.}\ \bibnamefont {Hensinger}},\ }\href {\doibase
  10.1103/PhysRevA.91.012322} {\bibfield  {journal} {\bibinfo  {journal} {Phys.
  Rev. A}\ }\textbf {\bibinfo {volume} {91}},\ \bibinfo {pages} {012322}
  (\bibinfo {year} {2015})}\BibitemShut {NoStop}%
\bibitem [{\citenamefont {Mikelsons}\ \emph {et~al.}(2015)\citenamefont
  {Mikelsons}, \citenamefont {Cohen}, \citenamefont {Retzker},\ and\
  \citenamefont {Plenio}}]{Gatis}%
  \BibitemOpen
  \bibfield  {author} {\bibinfo {author} {\bibfnamefont {G.}~\bibnamefont
  {Mikelsons}}, \bibinfo {author} {\bibfnamefont {I.}~\bibnamefont {Cohen}},
  \bibinfo {author} {\bibfnamefont {A.}~\bibnamefont {Retzker}}, \ and\
  \bibinfo {author} {\bibfnamefont {M.~B.}\ \bibnamefont {Plenio}},\ }\href
  {\doibase 10.1088/1367-2630/17/5/053032} {\bibfield  {journal} {\bibinfo
  {journal} {New J. Phys.}\ }\textbf {\bibinfo {volume} {17}},\ \bibinfo
  {pages} {053032} (\bibinfo {year} {2015})}\BibitemShut {NoStop}%
\bibitem [{\citenamefont {Schrieffer}\ and\ \citenamefont
  {Wolff}(1966)}]{polaron}%
  \BibitemOpen
  \bibfield  {author} {\bibinfo {author} {\bibfnamefont {J.}~\bibnamefont
  {Schrieffer}}\ and\ \bibinfo {author} {\bibfnamefont {P.}~\bibnamefont
  {Wolff}},\ }\href@noop {} {\bibfield  {journal} {\bibinfo  {journal}
  {Physical Review}\ }\textbf {\bibinfo {volume} {149}},\ \bibinfo {pages}
  {491} (\bibinfo {year} {1966})}\BibitemShut {NoStop}%
\bibitem [{\citenamefont {Aharon}\ \emph {et~al.}(2013)\citenamefont {Aharon},
  \citenamefont {Drewsen},\ and\ \citenamefont {Retzker}}]{nati}%
  \BibitemOpen
  \bibfield  {author} {\bibinfo {author} {\bibfnamefont {N.}~\bibnamefont
  {Aharon}}, \bibinfo {author} {\bibfnamefont {M.}~\bibnamefont {Drewsen}}, \
  and\ \bibinfo {author} {\bibfnamefont {A.}~\bibnamefont {Retzker}},\ }\href
  {\doibase 10.1103/PhysRevLett.111.230507} {\bibfield  {journal} {\bibinfo
  {journal} {Phys. Rev. Lett.}\ }\textbf {\bibinfo {volume} {111}},\ \bibinfo
  {pages} {230507} (\bibinfo {year} {2013})}\BibitemShut {NoStop}%
\bibitem [{\citenamefont {S{\o}rensen}\ and\ \citenamefont
  {M{\o}lmer}(2000)}]{Sorensen3}%
  \BibitemOpen
  \bibfield  {author} {\bibinfo {author} {\bibfnamefont {A.}~\bibnamefont
  {S{\o}rensen}}\ and\ \bibinfo {author} {\bibfnamefont {K.}~\bibnamefont
  {M{\o}lmer}},\ }\href@noop {} {\bibfield  {journal} {\bibinfo  {journal}
  {Phys. Rev. A}\ }\textbf {\bibinfo {volume} {62}},\ \bibinfo {pages} {022311}
  (\bibinfo {year} {2000})}\BibitemShut {NoStop}%
\bibitem [{\citenamefont {Roos}(2008)}]{Roos2}%
  \BibitemOpen
  \bibfield  {author} {\bibinfo {author} {\bibfnamefont {C.~F.}\ \bibnamefont
  {Roos}},\ }\href@noop {} {\bibfield  {journal} {\bibinfo  {journal} {New
  Journal of Physics}\ }\textbf {\bibinfo {volume} {10}},\ \bibinfo {pages}
  {013002} (\bibinfo {year} {2008})}\BibitemShut {NoStop}%
\bibitem [{\citenamefont {Deslauriers}\ \emph {et~al.}(2006)\citenamefont
  {Deslauriers}, \citenamefont {Olmschenk}, \citenamefont {Stick},
  \citenamefont {Hensinger}, \citenamefont {Sterk},\ and\ \citenamefont
  {Monroe}}]{Deslauriers1}%
  \BibitemOpen
  \bibfield  {author} {\bibinfo {author} {\bibfnamefont {L.}~\bibnamefont
  {Deslauriers}}, \bibinfo {author} {\bibfnamefont {S.}~\bibnamefont
  {Olmschenk}}, \bibinfo {author} {\bibfnamefont {D.}~\bibnamefont {Stick}},
  \bibinfo {author} {\bibfnamefont {W.~K.}\ \bibnamefont {Hensinger}}, \bibinfo
  {author} {\bibfnamefont {J.}~\bibnamefont {Sterk}}, \ and\ \bibinfo {author}
  {\bibfnamefont {C.}~\bibnamefont {Monroe}},\ }\href {\doibase
  10.1103/PhysRevLett.97.103007} {\bibfield  {journal} {\bibinfo  {journal}
  {Phys. Rev. Lett.}\ }\textbf {\bibinfo {volume} {97}},\ \bibinfo {pages}
  {103007} (\bibinfo {year} {2006})}\BibitemShut {NoStop}%
\bibitem [{\citenamefont {Piltz}\ \emph {et~al.}(2013)\citenamefont {Piltz},
  \citenamefont {Scharfenberger}, \citenamefont {Khromova}, \citenamefont
  {Var\'on},\ and\ \citenamefont {Wunderlich}}]{Piltz}%
  \BibitemOpen
  \bibfield  {author} {\bibinfo {author} {\bibfnamefont {C.}~\bibnamefont
  {Piltz}}, \bibinfo {author} {\bibfnamefont {B.}~\bibnamefont
  {Scharfenberger}}, \bibinfo {author} {\bibfnamefont {A.}~\bibnamefont
  {Khromova}}, \bibinfo {author} {\bibfnamefont {A.~F.}\ \bibnamefont
  {Var\'on}}, \ and\ \bibinfo {author} {\bibfnamefont {C.}~\bibnamefont
  {Wunderlich}},\ }\href {\doibase 10.1103/PhysRevLett.110.200501} {\bibfield
  {journal} {\bibinfo  {journal} {Phys. Rev. Lett.}\ }\textbf {\bibinfo
  {volume} {110}},\ \bibinfo {pages} {200501} (\bibinfo {year}
  {2013})}\BibitemShut {NoStop}%
\bibitem [{\citenamefont {Fowler}\ \emph {et~al.}(2012)\citenamefont {Fowler},
  \citenamefont {Mariantoni}, \citenamefont {Martinis},\ and\ \citenamefont
  {Cleland}}]{Fowler}%
  \BibitemOpen
  \bibfield  {author} {\bibinfo {author} {\bibfnamefont {A.~G.}\ \bibnamefont
  {Fowler}}, \bibinfo {author} {\bibfnamefont {M.}~\bibnamefont {Mariantoni}},
  \bibinfo {author} {\bibfnamefont {J.~M.}\ \bibnamefont {Martinis}}, \ and\
  \bibinfo {author} {\bibfnamefont {A.~N.}\ \bibnamefont {Cleland}},\ }\href
  {\doibase 10.1103/PhysRevA.86.032324} {\bibfield  {journal} {\bibinfo
  {journal} {Phys. Rev. A}\ }\textbf {\bibinfo {volume} {86}},\ \bibinfo
  {pages} {032324} (\bibinfo {year} {2012})}\BibitemShut {NoStop}%
\end{thebibliography}%

\clearpage

\end{document}